\documentclass[a4paper, 12pt]{article}

\usepackage[backend=bibtex,sorting=nyt,useprefix=false,style=authoryear]{biblatex}

\usepackage{geometry}
\usepackage[utf8]{inputenc}
\usepackage{setspace}
\usepackage{titlesec}
\usepackage{graphicx}
\usepackage{booktabs}
\usepackage{hyperref}
\usepackage{multirow} 
\usepackage{subcaption}
\usepackage{bm}
\usepackage{amsmath,amssymb,epsfig,amsthm,bm,xcolor,bm,enumerate,mathrsfs,graphics,caption}

\titleformat{\section}{\large\bfseries}{\thesection.}{0.5em}{\MakeUppercase}
\titleformat{\subsection}{\normalsize\bfseries}{\thesubsection.}{0.5em}{}

\newcommand{\R}{\mathbb{R}}

\newcommand{\p}{\mathbb{P}}

\newtheorem{theorem}{Theorem}

\bibliography{biblio}

\title{A mover-stayer model with time-dependent stayer fraction}
\author{
    \begin{tabular}{cc}
        Eni Musta & Martina Vittorietti \\
        University of Amsterdam & Delft University of Technology \\
        \texttt{e.musta@uva.nl} & \texttt{m.vittorietti@tudelft.nl}
    \end{tabular}
}
\date{}

\begin{document}

\maketitle
\begin{abstract}
	Mover-stayer models are used in social sciences and economics to model heterogeneous population dynamics in which some individuals never experience the event of interest  (``stayers"), while others transition between states over time (``movers"). Conventionally, the mover-stayer status is determined at baseline and time-dependent covariates are only incorporated  in the movers’ transition probabilities.  In this paper, we present a novel dynamic version of the mover-stayer model, allowing potential movers to become stayers over time based on time-varying circumstances. 
	{Using a multinomial logistic framework, our model incorporates both time-fixed and exogenous time-varying covariates to estimate transition probabilities among the states of potential movers, movers, and stayers. Both the initial state and transitions to the stayer state are treated as latent.} The introduction of this new model is motivated by the study of student mobility.  Specifically focusing on panel data on the inter-university mobility of Italian students, factors such as the students' change of course and university size are considered as time-varying covariates in modeling their probability of moving or becoming stayers; sex and age at enrollment as time-fixed covariates. We propose a maximum likelihood estimation approach and investigate {its finite-sample performance through simulations, comparing it to established models in the literature.} 
\end{abstract}

\section{Introduction}
To get insights on the dynamic of population phenomena, such as student mobility, labor market, or a particular disease, observation of their changes over time is necessary.
Many of the processes ruling these phenomena operate in continuous time, but for practical and economic reasons, subjects are often only observed at discrete time points (\cite{cook_mixed_1999}\, \cite{allison1982discrete}).
For example, time to moving to a different country, region or university for a student, could be viewed in principle as a continuous variable, whereas data of student mobility are generally available over a discrete time basis (e.g. in terms of academic years).
In fact, discrete-time-to-event analysis is often more appropriate in social and behavioral science research, where time is typically recorded in discrete intervals (\cite{muthen2005discrete}).
When subjects are assessed periodically over a time period, exact event or transition times are usually not observed, and only the state occupied at each assessment, together with the measurements of risk factors, is available. 
Under the most general scenario, the observation times may be regularly or irregularly spaced and may be unique to each subject. Here, we focus only on longitudinal studies where the observations are taken at regular time intervals, which are the same for all subjects, known as panel data. 

Markov chain models (\cite{anderson_statistical_1957}) are considered the state of the art to describe the population evolution among different states of interest as a function of time and of the probability of movement from state to state over time.
Standard time-homogeneous Markov chain models assume that everyone has the same probability of moving from state $i$ to state $j$, this movement depending only on individual being in state $i$. 
In applications of Markov models it is frequently assumed (often implicitly) that the population is homogeneous and therefore representable by a single Markov process (\cite{singer_social_1973}).
Since this assumption is often unrealistic, more general models have also been proposed, which include covariate information in modeling the transition probabilities. However, such models would still be overly simplistic and might lead to biased results in situations when not all subjects are susceptible (`at risk') for the event(s) of interest.

The mover-stayer model, firstly introduced by \cite{blumen_industrial_1955}, deals with this phenomenon.
It is a generalized Markov model 
where there are two latent types of individuals: movers and stayers. `Movers' are the subjects who have a positive probability of moving from one state to another over time and `stayers' will remain in their initial state throughout the observation period. 
The mover and stayer model has been extensively used to take into account the presence of unobserved heterogeneity{, in terms of susceptibility to the event of interest,} in several economic and demographic studies (see \cite{Frydman02092021} for creditworthiness of borrowers, \cite{fougere_bayesian_2003} and \cite{magidson2009using} for labour market and employment dynamics, \cite{saint2017movers} and \cite{cipollini2012firm} for farm and firm size dynamics, \cite{bijwaard_immigrant_2010}). 
Accounting for 
the presence of immune individuals, not at risk for the event of interest, is common also in other fields. 
In the medical field these models are known as
``cure models'' (\cite{boag_maximum_1949}), where the term `cured' corresponds to being not susceptible, and in the economics field as the ``split population models'' (\cite{schmidt1989predicting}).
These models have been employed to study a variety of phenomena, including the survival of cancer 
patients (\cite{legrand2019cure,musta2022short}), {perinatal epidemiology (\cite{stoltenberg2020cure})}, criminal recidivism {(\cite{de2022bayesian}), 
smartphone-based earthquake monitoring (\cite{aiello2025survival}). 

In the classical setting, the probability of being a stayer in a given state is the same for all individuals and is equal to the proportion of stayers in that state (\cite{goodman_statistical_1961}). In \cite{frydman_estimation_2018}, the authors propose an extension of the discrete-time mover-stayer model, in which the probability of being a stayer/mover depends on an individual's time-fixed covariates and is specified using the logistic model, which is a common choice also in the cure model setting. 
When data are collected over time, the values of some covariates can be time dependent 
and, in general, it is easy to incorporate them into discrete-time models (\cite{muthen2005discrete}). In the context of mover-stayer or cure models, 
various approaches have been developed for including time-dependent covariates in 
the transition probabilities matrix for the movers or in the probability of experiencing the event for the subjects at risk (see for example \cite{Dirick02012019}, \cite{beretta_variable_2019} and \cite{beger2017splitting}). 
However, the probability of being a stayer has been, {to the best of our knowledge,} 
always 
taken constant over time. This means that the mover-stayer or cure status is fixed and determined at baseline, which is often unrealistic.
For example, in the context of student mobility, one might initially be at risk of moving and then later become a stayer with a probability that depends on time-varying covariates {such as the achievement of the bachelor's degree.}

To allow for a stayer  status that can change over time, in this paper we propose a new version of the mover-stayer model with a time-dependent stayer fraction.
The idea is to take into account the initial heterogeneity of the population (stayers and potential movers) in combination with transition probabilities into the stayer or mover state. We use a logistic model with time-fixed covariates for the probability of initially being at risk or not and 
a multinomial logistic model with both time-fixed and exogenous time-varying covariates 
for the probability of becoming a mover or a stayer. 
{What makes the problem challenging is that, based on the observed censored data, we cannot distinguish between the individuals at risk and stayers. Particularly, transitions into the stayer state are not observed. Hence, parametric modeling assumptions are crucial for identifiability.}\\
The model proposed in this paper can be applied to several educational phenomena and more broadly to panel data settings with a proportion of stayers. The focus of this paper is student mobility {among universities}.
The causes of student mobility are numerous, both related to personal characteristics of the students, such as socioeconomic status, high school background and demographic factors and external characteristics, such as quality of the institution, territorial factors and peer effect (\cite{van2013determinants}).
The main assumption underlying all the literature, as far as the authors knowledge, is that every student is at risk of experiencing mobility.
However, there are students for whom mobility is not an option. A combination of socioeconomic status, territorial identity, high school background, and gender contributes to create the profile of the stayer student. Hence, in addition to estimating the probability of moving, it becomes crucial to better understand the stayer dynamics over time and how such probability depends on these underlying factors.

The paper is organized as follows.
After reviewing the standard mover and stayer model in Section \ref{sec:standardms}, we introduce the new version of the mover and stayer with time-dependent stayer status in Section \ref{sec:ourms}. In Section \ref{sec:simulation} we conduct a simulation study considering three settings with different sample size, length of the time intervals, and proportion of movers and compare our results with a model that does not take into account the presence of stayers and a model that considers it fixed in time. In Section \ref{sec:realdata}, an application to the mobility of Italian students {among different universities} is shown. Both time-fixed variables such as gender, high-school curriculum, macroregion of the university of enrollment, and age at enrollment and time-varying covariates such as the number of course changes 
and the size of the enrollment university are used.
Finally, in Section \ref{sec:conclusion}, results and future directions are discussed.
{The R-code for the implementation of the method can be found open-access at the following Github repository \url{https://github.com/eni-musta/Dynamic-mover-stayer-model}.}

\section{The standard Mover and Stayer Model}
\label{sec:standardms}
In the literature, the classical version of the mover and stayer model assumes that the movers will move following a Markov chain in discrete time. However, our setting considers the ``moving'' as the event of interest and therefore is more similar to the classical discrete-time cure setting (e.g. \cite{de_leonardis_default_2014}).
Consider a sample of $n$ independent individuals (students), who are being observed from a natural starting point $t=0$.
We assume that the population 
is a mixture of students at risk (potential movers) or not at risk (stayers). 
For a sample of size $n$, let $B_i$, $(i=1,\dots n)$ be a binary variable taking 
values $B_i=1$ if the $i-$th student is at risk and $B_i=0$ if not. Denote by $T_i$ the discrete random variable giving the time of event occurrence (i.e. the survival time).
Without loss of generality, we can assume 
that time takes only positive integer values ($t=0, 1, 2, 3, \dots $). In case time is inherently continuous, this can be seen as the discretized version with $T=t$ if the event happened during the interval $(t,t+1]$.
As in standard survival analysis, 
information for some students can be censored, meaning that the observation of the $i$-th subject continues until time $Y_i$, at which point either an event occurs ($Y_i=T_i$) or the observation is censored ($Y_i<T_i$).
For each subject $i$, let $\Delta_i$ be the censoring indicator denoting whether the student $i$ experiences the event of interest during the observation period ($\Delta_i=1$), or not ($\Delta_i=0$).
The censoring mechanism is supposed to be non-informative and conditionally independent of the stayer status and of the event time, given the covariates. 
Note that only three combinations of $B_i$ and $\Delta_i$ are possible: i) uncensored and at risk $B_i=1$ and $\Delta_i=1$; ii) censored and at risk $B_i=1$ and $\Delta_i=0$; iii) censored and not at risk $B_i=0$ and $\Delta_i=0$.

Let $\pi_i=\p(B_i=1{|\bm{x_i}})$ be the probability {for the $i$-th individual  being susceptible given a ${d}\times1$ vector of  time-fixed covariates $\bm{x}_{i}$.} 
{Let $\bm{z}_{it}$ denote a ${q}\times1$ vector of exogenous time-dependent covariates, i.e. their future values are not influenced by the occurrence of the event of interest (\cite{kalbfleisch2002statistical}). Let $P_{it}=\p(T_i=t|T_i\ge t,\bm{x_{i}},\bm{z_{it}}, B_i=1)$ denote the conditional probability for the $i$-th subject to experience the event of interest at time $t$, given that it is susceptible and that the event has not yet occurred. This is actually the discrete hazard function for the susceptible individuals and is usually assumed to depend only on the current value of the time-dependent covariates.}
The {overall} 
probability of {surviving beyond time $t$} 
can be expressed as:
\begin{equation}
{\p(T_i>t|\bm{x_{i}},\bm{\bar{z}_{it}})=1-\pi_i+\pi_i \prod_{s\leq t}(1-P_{is}),}
\label{eq:classicalms}
\end{equation}
{where $\bm{\bar{z}}_{it}=(\bm{z}_{is})_{s\leq t}$ denotes the whole past history of the covariate values.}

The next step is assuming a specific {model} 
for the discrete-hazard {function of the susceptible subjects and the initial stayer probabilities.} 
One of the most used choice is the logistic regression function:
\begin{equation}
\log[P_{it}/(1-P_{it})]=\bm{\beta}^{'}\bm{x_{i}}+\bm{\gamma}^{'}\bm{z_{it}},
\label{eq:logit}
\end{equation}
{where $\bm{\beta}$ and $\bm{\gamma}$ denote the unknown regression parameters corresponding to the fixed and time-dependent covariates.}
The main advantage of the logistic regression model is its computational convenience given by $P_{it}$ is constrained to lie in the unit interval. 
However, another common choice is the log-log complementary model, which yields very similar results for small time intervals (\cite{thompson1977treatment}).
Note that $\bm{\beta}$ contains an intercept {$\beta_{0{,t}}$}, which could depend on time but usually further restrictions are made to simplify estimation. For example, one can assume that $\beta_{0,t}$ is a polynomial function in $t$ with fixed degree (\cite{mantel1978lofistic,allison1982discrete}). Since terms depending on time can be incorporated in $\bm{z_{it}}$ as deterministic covariates, we assume that the intercept is constant $\beta_{0{,t}}=\beta_0$.} 
The initial probability of being susceptible is also usually estimated through a logit model:
\begin{equation}
{\pi_i=\pi(\alpha,\bm{x_i})}=\frac{\exp(\bm{x}_i^{'}\bm{\alpha})}{1+\exp(\bm{x}_i^{'}\bm{\alpha})},
\label{eq:curestatus}
\end{equation}
where $\bm{\alpha}$ is the vector of unknown parameters {containing an intercept term}. Note that, since the subjects are not presumed to {change their mover-stayer status} 
just time-fixed explanatory variables are considered.\\



\section{The Mover-Stayer model with a time-dependent stayer fraction }
\label{sec:ourms}
\begin{figure}[h]
\centering
\includegraphics[width=0.3\textwidth]{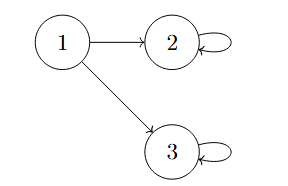}
\caption{State structure: State 1 - at risk of event (uncured, potential mover), State 2 - not at risk (cured, stayer), State 3 - observed event (relapse, mover)}
\label{fig:statestructure}
\end{figure}

In the classical mover and stayer setting, the probability of being susceptible or not is evaluated at initial time and kept constant over the whole time length.
To take into account the possibility that one subject can become stayer or cured {during the study period}, we propose a modification of the classical model. Let $S_t$ be the variable indicating the status of the subjects at time $t$, with state space $\mathcal{S}=\{1,2,3\}$, where `1' means at risk for the event of interest {(susceptible)}, `2' {means} not at risk {(immune)} and `3' {means event has occurred.} 
We assume {that initially the subjects can be in state 1 or 2 and that they} 
move according to the state structure in Figure \ref{fig:statestructure}. {Note that states 2 and 3 are absorbing states and transitions `1$\rightarrow$2' are not observed.}

{We consider panel data, meaning that transitions can happen at any time but subjects are only observed at regular intervals ($0, 1, 2, 3, . . . $). Let $\pi_i=\p(S_0=1|\bm{x}_i)$ be the probability of initially being  at risk given a vector of baseline covariates $\bm{x}_i$. Then, the {initial} probability of being a stayer 
is $\p(S_0=2|\bm{x}_i)=1-\pi_i.$ For $\pi_i$ we assume a logistic model as in \eqref{eq:curestatus}. Additionally, we define the probabilities of transitions `1$\rightarrow$2' and `1$\rightarrow$3' during the time interval $(t,t+1]$ as
\[
P_{i,1j}(t)=\p(S_{t+1}=j|S_t=1,\bm{x}_i,\bm{z}_{it}), \quad j=2,3,
\] 
where $\bm{z}_{it}$ again denotes a vector of exogenous time-dependent covariates. In particular, it is common to assume that the probability of a transition in $(t,t+1]$ depends only on the value of the time-dependent covariates at time $t$. For these probabilities, we use a parametric multinomial logistic model:}
\begin{equation}
\label{eqn:trans_prob}    \begin{aligned}
        P_{i,12}(t)&=\frac{\exp{(\bm{\gamma_{12}}^{'}\bm{z}_{it}+\bm{\beta_{12}}^{'}\bm{x}_i)}}{1+\exp{(\bm{\gamma_{12}}^{'}\bm{z}_{it}+\bm{\beta_{12}}^{'}\bm{x}_i)}+\exp{(\bm{\gamma_{13}}^{'}\bm{z}_{it}+\bm{\beta_{13}}^{'}\bm{x}_i)}}\\
        P_{i,13}(t)&=\frac{\exp{(\bm{\gamma_{13}}^{'}\bm{z}_{it}+\bm{\beta_{13}}^{'}\bm{x}_i)}}{1+\exp{(\bm{\gamma_{12}}^{'}\bm{z}_{it}+\bm{\beta_{12}}^{'}\bm{x}_i)}+\exp{(\bm{\gamma_{13}}^{'}\bm{z}_{it}+\bm{\beta_{13}}^{'}\bm{x}_i)}}\\
        P_{i,11}(t)&=1-P_{i,12}(t)-P_{i,13}(t),
    \end{aligned}
\end{equation}
where the parameters 
$\bm{\beta_{12}}$, $\bm{\beta_{13}}$  contain an intercept term  while $\bm{\gamma_{12}}$, $\bm{\gamma_{13}}$ no. We assume that the intercepts are constant {but, as mentioned in Section~\ref{sec:standardms}, one can also allow for a time dependent intercept (e.g. a polynomial function of time) by incorporating deterministic time-dependent covariates $t,t^2,t^3\dots$ within $\bm{z}_{it}.$}

{As in the standard model, the observation of the $i$-th subject continues until time $Y_i$, at which point either an event occurs ($Y_i=T_i$ meaning that the transition `1$\rightarrow$3' occurred in the interval $(Y_i,Y_i+1]$) or the observation is censored ($Y_i<T_i$).
For each subject $i$, let $\Delta_i$ be the censoring indicator denoting whether the student $i$ experiences the event of interest during the observation period ($\Delta_i=1$), or not ($\Delta_i=0$).}
When $\Delta_i=1$, we know that the individual was initially in state 1 and remained there until time $Y_i$, when he moves to state 3. When $\Delta_i=0$, there are 3 possible scenarios: the
individual was from the beginning in state 2; the individual started in state 1 and stayed there until time $Y_i$; the individual started in state 1 and moved to state 2 at some unknown time within $Y_i$. 
Censoring is assumed to be non-informative and conditionally independent of the transition times {and of the initial state, given the covariates.}

We can write the likelihood function, distinguishing  the contribution of censored and uncensored observations, as follows:
{\begin{equation}
\label{eqn:likelihood}
\begin{aligned}
    L(\bm{\alpha}, \bm{\beta}_{12},\bm\beta_{13}, \bm{\gamma}_{12},\bm\gamma_{13}) &= \left(\pi_i\prod_{t=0}^{Y_i-1}P_{i,11}(t)P_{i,13}(Y_i)\right)^{\Delta_i}\\
&\qquad\times\left(1-\pi_i+\pi_i\prod_{t=0}^{Y_i}P_{i,11}(t)+\pi_i\sum_{s=0}^{Y_i}\prod_{t=0}^{s-1}P_{i,11}(t)P_{i,12}(s)\right)^{1-\Delta_i}
    \end{aligned}
\end{equation}
with the convention that an empty product is equal to 1.}
{Note that we are assuming the time-dependent covariates are exogenous and particularly
\[
\p(\bm{z}_t=\bm{z}|\bm{x},\bm{\bar{z}}_{t-1},S_{t-1}=1)=\p(\bm{z}_t=\bm{z}|\bm{x},\bm{\bar{z}}_{t-1},S_{t-1}=2),
\]
i.e. the distribution of future values of $\bm{z}_t$ is the same for individuals in state 1 and 2. Otherwise, covariates would provide information to distinguish between the three possible latent scenarios for the censored observations. As a result, the likelihood function (for censored observations) would also contain terms corresponding to the distribution of the covariates for each state, requiring a joint modeling approach.} 

Estimation of the model parameters can be done via maximum likelihood. One can compute the maximum likelihood estimators via direct optimization of the likelihood function over the finite-dimensional parameters or using the EM-algorithm. {Details of the EM algorithm are provided in Appendix~\ref{sec:EM}.} 
 {In this paper, we opt for direct likelihood maximization over all parameters and, to avoid problems with local maxima, the procedure is repeated several times with different random initializations.}

Based on standard asymptotic theory for the maximum likelihood estimators, since the likelihood function is continuously differentiable, we have the following result.
\begin{theorem}
\label{theo:normality}
Let $\hat{\bm\theta}_n=(\hat{\bm\alpha},\hat{\bm\beta}_{12},\hat{\bm\beta}_{13},\hat{\bm\gamma}_{12},\hat{\bm\gamma}_{13})$ denote the maximum likelihood estimator of the true parameter $\bm\theta=(\bm\alpha,\bm\beta_{12},\bm\beta_{13},\bm\gamma_{12},\bm\gamma_{13})$. Assume that $\theta$ is an interior point of $\Theta$, for some compact and convex subset $\Theta$ of $\R^{3(d+1)+2q}$, and that the support of the covariates is bounded. Then, as $n\to\infty$,
\[
\sqrt{n}(\hat{\bm\theta}_n-\bm\theta)\xrightarrow{d} N(0,I_\theta^{-1}), 
\]
where $I_\theta$ denotes the Fisher information matrix.
\end{theorem}
\noindent Note that for Theorem~\ref{theo:normality} to be valid, the model needs to be identifiable, meaning that
\[ L(\bm\theta|Y,\Delta,\bm{x},\bar{\bm{z}}_{Y})= L(\tilde{\bm\theta}|Y,\Delta,\bm{x},\bar{\bm{z}}_{Y}) \text{ almost surely} \Rightarrow \bm\theta=\tilde{\bm\theta}.
\]
In our setting, identifiability is ensured thanks to the parametric multinomial logistic assumptions on the transition probabilities and the logistic model for the initial state, under the condition that the covariates are not multicollinear. Note that identifiability for a nonparametric or semiparametric modeling approach would not be possible since the transitions into state 2 are not observed and as a result one cannot distinguish between those who are in state 1 or 2. Despite the model being theoretically identifiable, for moderate sample sizes one can still encounter practical identifiability problems, i.e. there are several values of the parameters for which the likelihood is maximized.  This can be detected by performing the likelihood maximization several times starting from different initial points. 
Another issue that can arise in practice is separation as in the standard logistic regression models, meaning that the likelihood is maximized when certain probabilities are 0 or 1, leading to non-existence of the MLE for the regression parameters (\cite{albert1984existence}). When this phenomenon occurs, we observe estimates that are extremely large and the likelihood keeps increasing as the absolute value of the parameters increases. A potential remedy in such scenarios is to use a penalized likelihood approach, which stabilizes the parameter estimates at the cost of  introducing bias. 
 
 Using Theorem~\ref{theo:normality}, one could construct asymptotic confidence intervals for the parameters of interest. However, the asymptotic covariance matrix is not easy to compute analytically because of the complex expression of the likelihood function. One could estimate the asymptotic variance via the numerical approximation of the Hessian matrix or use a bootstrap procedure. The practical performance of both approaches is investigated in Section \ref{sec:simulation}.
 
The obtained estimates allow also compute the cumulative probabilities of moving or becoming a stayer for $t\geq 1$:
 \begin{equation}
\label{eqn:cum_prob}
\begin{aligned}
    P_{i,2}(t)&=\p(S_{t}=2\mid \bm{\bar{z}_{it}},\bm{x}_i)=\{1-\pi_i\}+\pi_i\sum_{s=0}^{t-1}\prod_{u=0}^{s-1}P_{i,11}(u)P_{i,12}(s)\\
    P_{i,3}(t)&=\p(S_{t}=3\mid \bm{\bar{z}_{it}},\bm{x}_i)=\pi_i\sum_{s=0}^{t-1}\prod_{u=0}^{s-1}P_{i,11}(u)P_{i,13}(s).
\end{aligned}
\end{equation}
These probabilities can be used to make future predictions for `typical' trajectories of the external covariates $\bm{z}_t$. 

\section{Simulation study}
\label{sec:simulation}
In this section we investigate the finite sample performance of the method through a simulation study. To illustrate the advantages of the proposed dynamic model we also compare with the standard (static) mover-stayer model, which assumes that the stayer status is fixed and determined at baseline, and with a standard {discrete} time-to-event model which accounts for time-varying covariates but ignores the possibility of being a stayer. 
We consider three different simulation settings, one of which is presented here, while the other two can be found in Appendix~\ref{sec:appendix}. 

\textbf{Setting 1.} We consider two independent baseline covariates {$\bm{x_i}=(x_i^1,x_i^2)$}: $x_i^1$ has a standard normal distribution and $x_i^2$ is a Bernoulli random variable with success probability $0.4$. The baseline mover-stayer status $B_0$ is generated as a Bernoulli variable with success probability $\pi(\bm\alpha;\bm{x}_i)$ given in \eqref{eq:curestatus} with $\bm\alpha=(0.8,0.5,-1)$.  We consider a maximum length follow-up $K=5$ and the event (moving) time $T$ can take values in $\{0,1,2,3,4\}$. At the beginning of each time unit we observe a bivariate vector of external time-dependent covariates {$\bm{z}_{it}$} with components defined as follows. 
{\[
\begin{cases}
 z^1_{i0}&=0 \\
 z^1_{it}&=z^1_{it-1}+U^1_{it}
\end{cases}\qquad \begin{cases}
    z^2_{i0}&\sim \text{Uniform}(\{1,2,3,4,5\}) \\
 z^2_{it}&=z^2_{it-1}+U^2_{it-1}
\end{cases}
\]}
where the variables $(U^1_t,U^2_t)_{t\in\{0,\dots,4\}}$ are all independent and identically distributed, $U^1_t\sim N(0.5,1)$ and {$U^2_t\sim \text{Binom}(2,0.5)$}. At each time, we update the mover-stayer status $S_t$ according to the transition probabilities given in~\eqref{eqn:trans_prob} conditional on the covariate values with $\bm\beta_{12}=(-1,0.6,-0.1)$, $\bm\beta_{13}=(-2,-0.4,0.1)$, $\bm\gamma_{12}=(0.11,-0.2)$ and $\bm\gamma_{13}=(-0.5,0.3)$.  The censoring times are first generated from the continuous shifted exponential distribution with parameter 0.03 on $(1,\infty)$, and then are discretized and truncated at $4$. Table~\ref{tab:setting1} shows, for each time point, the percentage of observations in each of the three states, the percentage of observed movers and of censored observations. 

\begin{table}[h!]
    \centering
    \begin{tabular}{c|rrr|r|r}
    Time & in state 1 & in state 2& in state 3 & Obs. movers & Censored\\
    \hline
    0 & 59\% & 41\%& 0\%  & 13.2\%  & 0.0\%  \\
    1 & 36\% & 51\% & 13\%  &  6.6\% & 3.4\% \\
    2 & 23\% & 57\%&20\%  &3.4\%   &3.0\%  \\
    3 & 15\%&61\% & 24\% &  1.8\% & 2.7\% \\
    4 &10\% &64\% &26\%  &  1.0\% & 64.5\% \\
    5 & 7\% & 66\%& 27\% &   &  \\
    \end{tabular}
    \caption{Percentage of observations in each state, of observed movers and of censored observations for each time point in Setting 1.}
    \label{tab:setting1}
\end{table}

In this setting the initial proportion of stayers is quite high and about $25\%$ of the subjects who are initially at risk become stayers during the study. As a result there is only a small proportion of transitions to the mover state. Note also that we have light censoring during the study and most of the observations are censored at the last time point. Particularly, all the stayers are observed as censored since the transition to the stayer state is latent. 

\begin{figure}[h!]
    \centering
\includegraphics[width=0.8\textwidth]{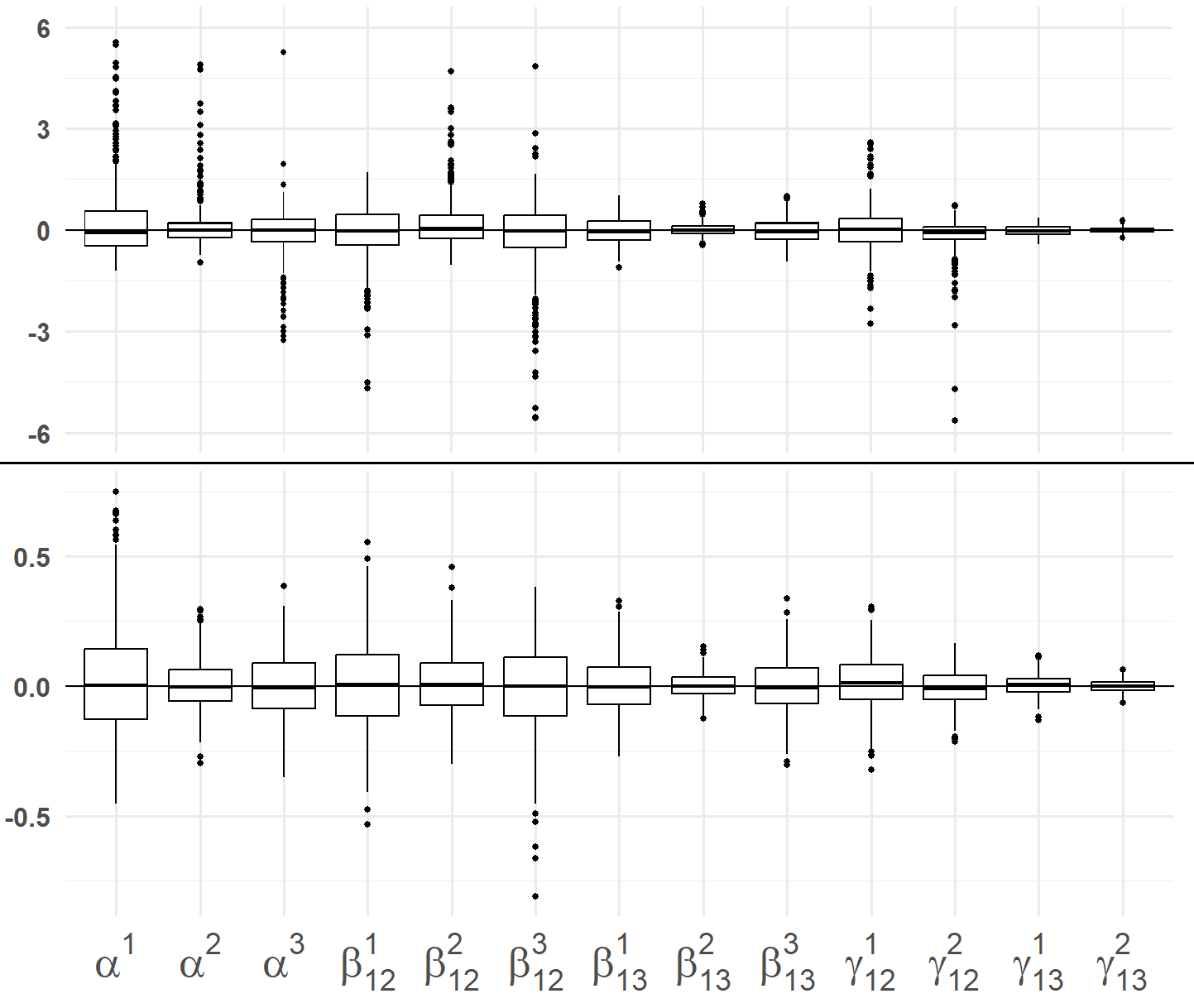}
    \caption{Boxplots of the parameter estimates centered at the true value for Setting 1 and sample size $n=1000$ (top), $n=10000$ (bottom).}
    \label{fig:est_1}
\end{figure}

We consider two sample sizes $n=1000$ and $n=10000$ and for each sample size generate 500 replications of the data. Maximum likelihood estimates of the parameters are computed using the \texttt{optim} function in \texttt{R} with the true parameter values as initial point. The computational time is about 2-3 min per iteration for $n=1000$ and about 15 min per iteration for $n=10000$, with  a processor 13th Gen Intel(R) Core(TM) i7-1355U   1.70 GHz. In practice the true parameters are not known and several random initialization would usually be required to guarantee identification of the global maximum. However, to reduce the computation cost for the purpose of the simulation study, we chose to just perform maximization once with an `optimal' starting point, {i.e. the true parameters}. Boxplots of the parameter estimates over the 500 replications,  centered at the true values (i.e. $\hat{\bm\theta}_n-\bm\theta_0$ where $\bm\theta_0$ denotes the vector of the true parameters and $\hat{\bm\theta}_n$ its maximum likelihood estimate),  are shown in Figure~\ref{fig:est_1}. We observe that the boxplots are centered at zero, as they should, and the variability decreases as the sample size increases. As expected, estimation of the parameters related to the transition to the mover state is more accurate, since these transitions are observed. The parameters related to the initial status and the transitions to the stayer state, which are latent, have higher variability. In particular, for sample size 1000, {unrealistically large estimates} are sometimes observed.  
However, in order to maintain good visibility of the figure, estimates with absolute value exceeding 6 are not shown and that occurs in about $5\%$ of cases. Such extreme estimates do not seem to be caused by local maxima but rather indicate problems with practical identifiability as a result of a `separation' phenomenon, as mentioned in Section~\ref{sec:ourms}. {In practice}, one would be able to detect if this problem occurs since the likelihood would keep increasing as the parameters explode. A possible solution in such cases would be to consider  penalized maximum likelihood estimation, which would decrease variability but also introduce a slight bias.
{However, we do not explore this direction further here due to challenges related to selecting an appropriate penalty parameter and the complexities of performing valid inference after model selection.} 
 When the sample size increases, estimation becomes stable and we did not encounter any identifiability problem. 

\begin{figure}[h!]
    \centering
    \includegraphics[width=0.8\textwidth]{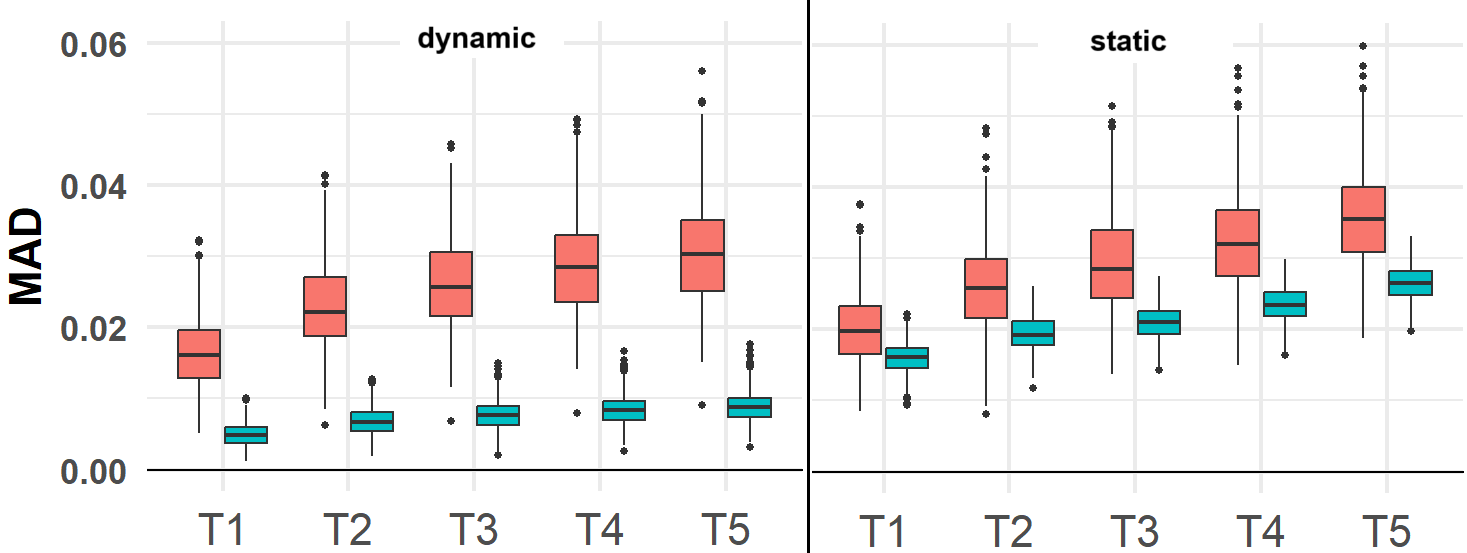}\\    \includegraphics[width=0.45\textwidth]{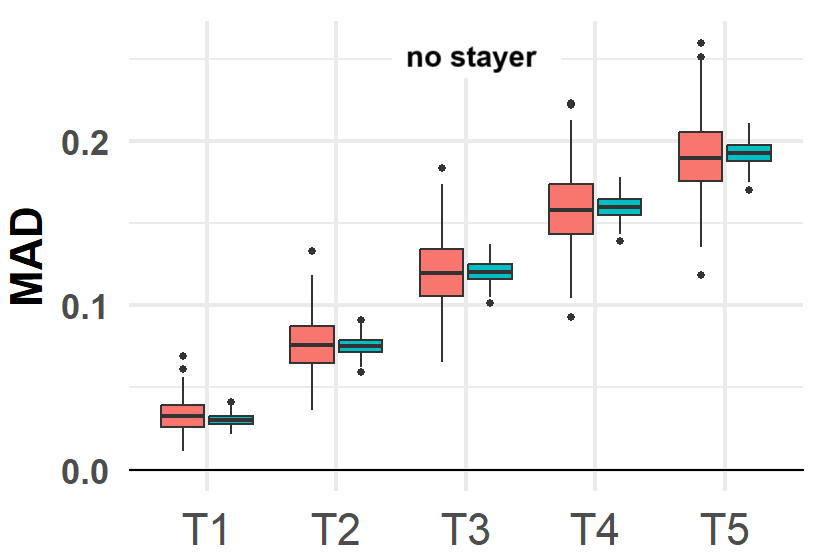}\\
    \caption{Boxplots of the average mean absolute deviation (MAD) for the cumulative probabilities of being a mover in Setting 1 and sample size $n=1000$ (red), $n=10000$ (green). Top left: dynamic model; Top right: static model; Bottom: model without stayers. }
    \label{fig:MAD_mover_1}
\end{figure}

\begin{figure}[h!]
    \centering
    \includegraphics[width=0.8\textwidth]{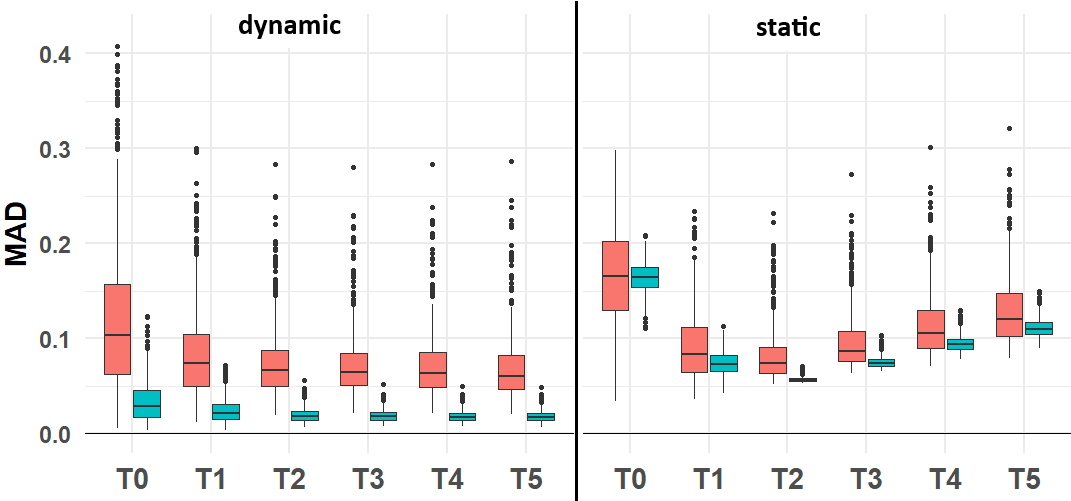}
    \caption{Boxplots of the average mean absolute deviation (MAD) for the cumulative probabilities of being a stayer in Setting 1 and sample size $n=1000$ (red), $n=10000$ (green). Left: dynamic model; Right: static model.}
    \label{fig:MAD_stayer_1}
\end{figure}

In addition to the parameter estimates, we also evaluate the accuracy of the estimated probabilities. For $t\in\{0,1,\dots,{5}\}$, we compute the time-dependent average mean absolute deviation of cumulative probabilities of being a mover (k=3) or a stayer (k=2) defined as
\begin{equation*}
    \text{MAD}_k(t)=\sum_{i=1}^n \left|\p(S_{t}=k|\bm{x}_i,\bar{\bm{z}}_{it};\bm\theta_0)-\p(S_{t}=k|\bm{x}_{{i}},\bar{\bm{z}}_{it};\hat{\bm\theta}_n)  \right|.
\end{equation*}
Note that MAD$_2(0)$ corresponds to the average mean absolute deviation of the baseline probabilities of being a stayer. On the other hand, for $k=3$, only MAD at $t\in\{1,\dots,{5}\}$ is relevant since MAD$_3(0)=0$. The boxplots of the mean absolute deviation of the mover and stayer probabilities are shown in Figures~\ref{fig:MAD_mover_1}-\ref{fig:MAD_stayer_1} respectively. As expected, the probabilities of being a mover are estimated more accurately than the probabilities of being a stayer. As the sample size increases, the MAD becomes smaller and has less variability. 

In addition to the proposed dynamic model, we also fit a static model and a model without stayers. In both cases, we use a logistic model as in \eqref{eq:logit} for the event probability  at time $t$ (in the static mover-stayer model only for the individuals at risk).  We fitted both models with constant intercept and time-varying intercept given by a polynomial function of time with degree 3. Here we just present the results for the constant intercept case since the more general model with time-varying intercept did not lead to better performance. 
{We compare our model with these two models only in terms of MAD  of the estimated cumulative probabilities given that their parameter estimates are misspecified.} Note that for the static model, the stayer probabilities are constant over time, while the mover probabilities correspond to the cumulative distribution function of the event times. The model without stayers allows us to only estimate the cumulative probabilities of being a mover. From the boxplots in Figures~\ref{fig:MAD_mover_1}-\ref{fig:MAD_stayer_1} we see that ignoring the existence of stayers, who are not at risk for the event of interest, clearly leads to biased estimates for the mover probabilities, and this bias does not get smaller as the sample size increases. For $n=1000$, the static model is comparable to the dynamic model when estimating mover probabilities and slightly worse for stayer probabilities. However, when $n=10000$, the dynamic model performs significantly better. 


Such conclusions remain valid also for the other two simulation settings presented in Appendix~\ref{sec:appendix}. For setting 2, which has a low initial proportion of stayers and a larger proportion of observed movers, the static model performs slightly better than the dynamic model for estimating mover probabilities when $n=1000$. However, as $n$ increases, the dynamic model improves significantly and has the lowest MAD. For setting 3, which has a longer maximum follow-up of $K=10$, parameter estimates are more precise and the mean absolute deviation of both mover and stayer probabilities is lower. The dynamic model performs the best, particularly for large sample size. In all scenarios, the model without stayers gives biased results.

 Finally, we also investigated the construction of confidence intervals for the unknown parameters. Asymptotic Wald confidence intervals were constructed using a numerical estimate of the Fisher information matrix (via the Hessian matrix computed from the \texttt{optim} function), since explicit calculation is not simple {due to} the form of the likelihood function. However, we observe that this method underestimates the standard errors leading to confidence intervals with lower coverage than the nominal level of $95\%$ even when the sample size is large (see Table~\ref{tab:CI} for Setting 1). This is probably due to numerical instability in computation of the Hessian matrix and its inverse. Hence, we recommend the use of Wald confidence intervals with bootstrap estimates of the standard errors. For $n=1000$, bootstrap is still problematic since extreme values of the estimates for some of the bootstrap samples inflate the standard errors leading to unrealistically {large} 
 confidence intervals, which are also very sensitive to the number of bootstrap samples. For $n=10000$, we expect good performance of the bootstrap confidence intervals but a study of the coverage probabilities is infeasible due to the computational cost. The warp speed bootstrap (\cite{giacomini2013warp}) has been proposed to approximate the coverage probabilities by generating just one, instead of a large number, of bootstrap sample for each replicated dataset. However, this procedure is known to give a higher coverage than the actual one  (because of overestimation of the standard errors) and we indeed observe from the results in Table~\ref{tab:CI} that the coverage is almost one. To get an idea of the actual bootstrap estimates of the standard errors, we perform the bootstrap procedure with 200 bootstrap samples for 5 random samples generated from Setting 1. Results in Table~\ref{tab:sd_boot} show that the bootstrap estimates are closer to the actual standard errors computed over the 500 replicated datasets. This suggests that the Wald confidence intervals with bootstrap estimates of the standard errors have a good performance in practice. 
 \begin{table}[h!]
    \centering
    \begin{tabular}{c|ccc|ccc}
    & \multicolumn{3}{c|}{M1} & \multicolumn{3}{c}{M2}  \\
    Parameter & Cov & Length & sd & Cov & Length & sd\\
    \hline
    $\alpha[1]$  & 0.89  & 0.67  & 0.171 & 0.998 & 1.26  & 0.321 \\
       $\alpha[2]$  & 0.83  & 0.26  & 0.067 & 0.994 & 0.57  & 0.144 \\
       $\alpha[3]$  & 0.92  & 0.44  & 0.111 & 0.992 & 0.78  & 0.200 \\
       $\beta_{12}[1]$  & 0.90  & 0.66  & 0.167 & 0.994 & 1.02  & 0.261 \\
       $\beta_{12}[2]$  & 0.91  & 0.39  & 0.099 & 0.998 & 0.72  & 0.184 \\
       $\beta_{12}[3]$  & 0.94  & 0.67  & 0.172 & 0.990 & 1.04  & 0.265 \\
       $\beta_{13}[1]$  & 0.87  & 0.34  & 0.088 & 0.996 & 0.63  & 0.161 \\
       $\beta_{13}[2]$  & 0.91  & 0.16  & 0.042 & 0.998 & 0.27  & 0.068 \\
       $\beta_{13}[3]$  & 0.91  & 0.35  & 0.090 & 0.986 & 0.59  & 0.151 \\
       $\gamma_{12}[1]$  & 0.93  & 0.40  & 0.103 & 0.998 & 0.69  & 0.175 \\
       $\gamma_{12}[2]$  & 0.72  & 0.15  & 0.038 & 0.994 & 0.40  & 0.101 \\
       $\gamma_{13}[1]$  & 0.89  & 0.14  & 0.035 & 0.994 & 0.25  & 0.063 \\
       $\gamma_{13}[2]$  & 0.72  & 0.05  & 0.013 & 0.998 & 0.15  & 0.038 \\
    \end{tabular}
    \caption{Coverage probability and average length of 95\% Wald confidence intervals for the parameters in Setting 1 and sample size $n=10000$. The standard error of the estimates (sd) is estimated using the numerical Hessian matrix of the likelihood function (M1) or the warp speed bootstrap (M2).}
    \label{tab:CI}
\end{table}
\begin{table}[h!]
    \centering
    \begin{tabular}{c|ccccc|c}
     & \multicolumn{5}{c|}{Bootstrap SD estimate} &SD \\
    Parameter & Sample 1 & Sample 2&  Sample 3& Sample 4 & Sample 5 &  (500 replications)\\
    \hline
     &  & & &&&\\[-10pt]
       $\alpha[1]$  & 0.215 & 0.198 & 0.176 & 0.205 & 0.167&0.209\\
         &  & & &&&\\[-10pt]
       $\alpha[2]$  & 0.097 & 0.094 & 0.079 & 0.088 & 0.077&0.095\\
         &  & & &&&\\[-10pt]
       $\alpha[3]$  &0.136 & 0.130 & 0.125 & 0.141 & 0.125 &0.124\\
         &  & & &&&\\[-10pt]
       $\beta_{12}[1]$  &0.147 & 0.168 & 0.177 & 0.207 & 0.243&0.173\\
         &  & & &&\\[-10pt]
       $\beta_{12}[2]$  & 0.109 & 0.115 & 0.139 & 0.115 & 0.239&0.122\\
         &  & & &&&\\[-10pt]
       $\beta_{12}[3]$  & 0.154 & 0.159 & 0.175 & 0.188 & 0.312& 0.177\\
         &  & & &&&\\[-10pt]
       $\beta_{13}[1]$  & 0.104 & 0.106 & 0.108 & 0.122 & 0.093&0.107 \\
         &  & &&&&\\[-10pt]
       $\beta_{13}[2]$  & 0.045 & 0.043 & 0.048 & 0.047 & 0.047 &0.046\\
         &  & && &&\\[-10pt]
       $\beta_{13}[3]$  & 0.097 & 0.099 & 0.103 & 0.097 & 0.103&0.103\\
         &  & &&& &\\[-10pt]
       $\gamma_{12}[1]$  & 0.117 & 0.107 & 0.121 & 0.131 & 0.172&0.110\\
         &  & & &&&\\[-10pt]
       $\gamma_{12}[2]$  & 0.060 & 0.060 & 0.069 & 0.091 & 0.092&0.063\\
         &  & &&& &\\[-10pt]
       $\gamma_{13}[1]$  &0.048 & 0.038 & 0.047 & 0.046 & 0.042& 0.042\\
         &  & && &&\\[-10pt]
       $\gamma_{13}[2]$  & 0.024 & 0.023 & 0.026 & 0.026 & 0.023&0.024\\
         &  & & &&&\\[-10pt]
    \end{tabular}
    \caption{Standard errors of the parameter estimates for 5 random samples from Setting 1 and sample size $n=10000$, estimated using 200 bootstrap samples, compared with standard errors of the estimates computed over 500 replications.}
    \label{tab:sd_boot}
\end{table}

\clearpage
\section{Real Data example}
\label{sec:realdata}
The dataset used in this paper is a sample from the aggregate longitudinal database MOBYSU.IT (\cite{MOBISU2022}) combining cross-sectional data from the Italian Ministry of Education and Research, the National Student Registry, and/or from internal archive of specific universities. In this
paper, the focus is on Italian students that enrolled in an Italian University in 2014 longitudinally followed up until 2019 (30000 records in total). Italy with its geographical gap, the common North–South divide, and its experience of brain drain constitutes an interesting case in student mobility analysis (\cite{vittorietti2023new}).

More specifically, we are interested in the student mobility flows that occur after the enrollment, over the Italian territory, from university to university. 
The influence of several time-fixed covariates on student mobility among different provinces and macro-geographical areas have been studied in the literature (\cite{attanasio2020verso}).
The macroregion of origin is one of the main risk factors. Student mobility in Italy primarily flows in one direction, from the South to the Center and North (\cite{attanasio2022students}).
Also the high-school background plays an important role: students having a high school diploma in a scientific or humanistic area tend to have a higher mobility attitude with respect to the fellows that have a vocational or technical high school diploma (\cite{d2019out}). The high-school diploma in Italy is in fact strictly related to the socioeconomic status of the family and in \cite{d2019out} and \cite{rizzi2021moving} it was found that classical and scientific ``Liceo'' students are generally middle-upper class, and, thus, they are more likely to migrate.
There is also gender bias in student mobility: \cite{sulis2024gender} found a lower mobility attitude of female students compared to male students from the South and Islands.
Having a Bachelor’s or Master’s degree highly influences the mobility attitude of
the students, confirming the conjecture that education plays a very important role
in the mobility of the Italian students (\cite{iammarino2015education}).
\cite{de2017new} investigate the role of university quality indicators on Italian student mobility. Students may now pay more attention to the quality and
effectiveness of their tertiary education experience, partly to improve their chances of finding better employment conditions after graduation. The size of the university and a positive increasing trend in the number of enrollments are two of the major indices in several Italian university rankings (\cite{lun2006stability}).
\cite{Ciriaci03102014} considered the university size categorized into four categories: small, medium, large and mega as a possible explanatory variable for the probability of moving and found out that having graduated from `mega' university increases the probability to migrate after graduation, suggesting the idea that smaller universities may offer relatively higher teaching standards and/
or more efficient placement services.

\begin{figure}[h!]
   \centering
   \includegraphics[width=\textwidth]{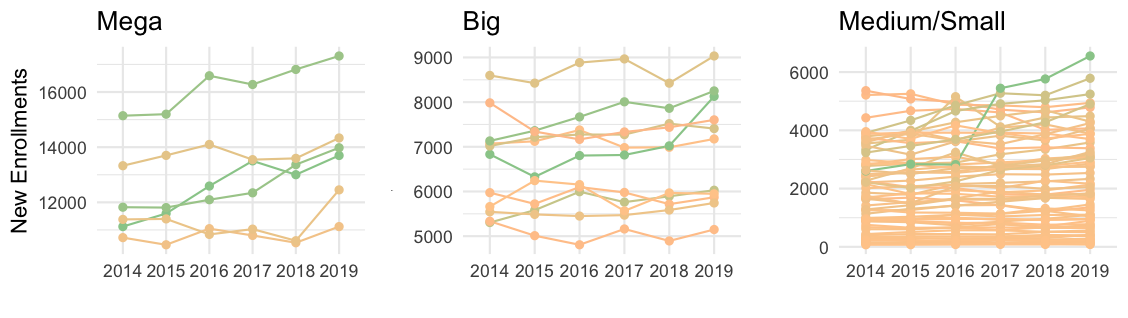}
    \caption{New enrollments in Mega, Big and Medium-Small Italian Universities from 2014 to 2019.}
    \label{fig:size}
\end{figure}

Hence, we will use as time-fixed covariates: sex, age at enrollment, high-school diploma {(Lyceum vs Non-Lyceum)} and macroregion at enrollment (North, Center, South and Islands). We will include as time-varying covariates the number of new enrollments of the university of origin and a counter variable that counts the number of course changes a student performs.\\
Censoring in this context happens if a student dropout. Student dropout is a complex phenomenon and it can be itself an event to study or as in \cite{enea2018south} can be considered as a competing risk event with graduation. However, we consider a relatively short time interval (5 years), so we can assume censoring is just attributable to dropout and not to graduation.\\
Table \ref{tab:realdata1} shows the time-fixed variables to be used in the model.
Our sample consists mainly of female students (54.7\%), enrolled in a Northern university (49.3\%) with an average age of 19.9 and who have attended a Lyceum as high school (58.5\%).
\begin{table}[h!]
\centering
\begin{tabular}{l|c|c}
{Variable} & {Mean} &{Standard Deviation} \\
\hline
Diploma (Lyceum) & 0.5849 & - \\
Sex (Male)     & 0.4532 & - \\
North   & 0.4934 & - \\
South   & 0.2730 & - \\
Age at enrollment     & 19.8607 & 3.2629 \\
\end{tabular}
\caption{Sample Means and Standard deviation of the time-fixed covariates ($n=30000$). }
\label{tab:realdata1}
\end{table}
In Table \ref{tab:realdata2} the percentage of observed movers and censored observations for each time point is reported. It is interesting to notice the drop in observed movers at time 1. This is reasonable since it corresponds to the time in which students settle. The highest probability is observed at time 3, which could correspond to the year of the Bachelor's graduation. 
\begin{table}[h!]
    \centering
    \begin{tabular}{c|r|r}
    Time & Obs. movers ($n=4940$) & Censored ($n=25060$)\\
    \hline
    0  & 20.1\%  & 7.6\%  \\
    1  &  8.2\% & 2.5\% \\
    2   &28.7\%   &2.3\%  \\
    3  &  29.5\% & 2.2\% \\
    4 &  13.6\% & 85.4\% \\
    \end{tabular}
    \caption{Percentage of observed movers and censored observations with respect to the total number of movers and censored respectively, for each time point ($n=30000$).}
    \label{tab:realdata2}
\end{table}
In Figure \ref{fig:size} and Table \ref{tab:percentage_distribution} the distribution of the time-varying covariates is shown.
In particular, from Figure \ref{fig:size} it is worth to note that the trend of new enrollments is predominantly constant in the medium/small universities and instead among the mega ans big universities is more common to find positive increasing trends.
Table \ref{tab:percentage_distribution} shows that 57.7\% of students in our sample have performed at least one change of course from 2014 to 2019. It is important to recall that the Bachelor's graduation is also included as a change of course, which is why a higher percentage of change course is seen after 3 years. 
\begin{table}[h!]
\centering
\begin{tabular}{l|c|c|c|c|c}
{Year} & {0} & {1} & {2} & {3} & {4} \\
\hline
2014 & 100.0 &  &  &  &  \\
2015 & 91.9  & 8.1   &  &  &  \\
2016 & 88.0  & 10.7  & 0.3  &  &  \\
2017 & 66.9  & 32.5  & 0.9  & 0.01 &  \\
2018 & 49.3  & 47.3  & 2.4  & 0.03 &  \\
2019 & 42.3  & 54.1  & 4.0  & 0.1  & 0.01 \\
\end{tabular}
\caption{Percentage distribution (by row) of the student by change course variable for each year ($n=30000$)}
\label{tab:percentage_distribution}
\end{table}

\subsection{Results}
In Table \ref{tab:bootstrap_comparison} the results for the dynamic and static model, previously introduced in section \ref{sec:simulation} are shown.
Both models are fitted with a cubic polynomial time-varying intercept, because of better performance in terms of AIC with respect to the static and linear/quadratic intercept.
The AIC of the dynamic model is 393391.92 lower than the one of  the static model, 39612.48, suggesting to use a time-dependent stayer fraction.

As for the initial probability of being at risk of moving, having attended a lyceum increases the probability of being susceptible to the event ($\alpha$[Lyceum]$=12.2$) (Table \ref{tab:bootstrap_comparison}). The strong effect of the high school background is visible also on the estimated mean probabilities of being at risk shown in Table \ref{tab:summary_probs}. 
Being older at enrollment and male increases the probability of becoming stayer with respect to remaining at risk ($\beta_{12}$[Age]$=0.430$, $\beta_{12}$[Male]$=0.126$), instead being enrolled in a southern universities decreases the probability of becoming a stayer with respect to the students enrolled in a university in the center of Italy ($\beta_{12}$[South]$=-0.370$).\\
As for the probability of moving with respect to the probability of remaining at risk of moving, being older and from the south are protective factors ($\beta_{13}$[Age]$=-0.200$, $\beta_{14}$[South]$=-0.173$), instead it is more probable to experience the event for students from the north ($\beta_{13}$[North]$=0.178$).
The size of the university increases the probability of becoming a stayer and decreases the probability of moving ($\gamma_{12}$[Size]$=0.080$, $\gamma_{13}$[Size]$=-0.080$); on the other hand, the higher  the number of changes of course a student experiences, the more probable it is to remain at risk of moving ($\gamma_{12}$[Change]$=-17.000$, $\gamma_{13}$[Change]$=-3.150$).

\begin{figure}[htbp]
    \centering
    
    \begin{subfigure}[b]{0.50\textwidth}
        \centering
        \includegraphics[width=\textwidth]{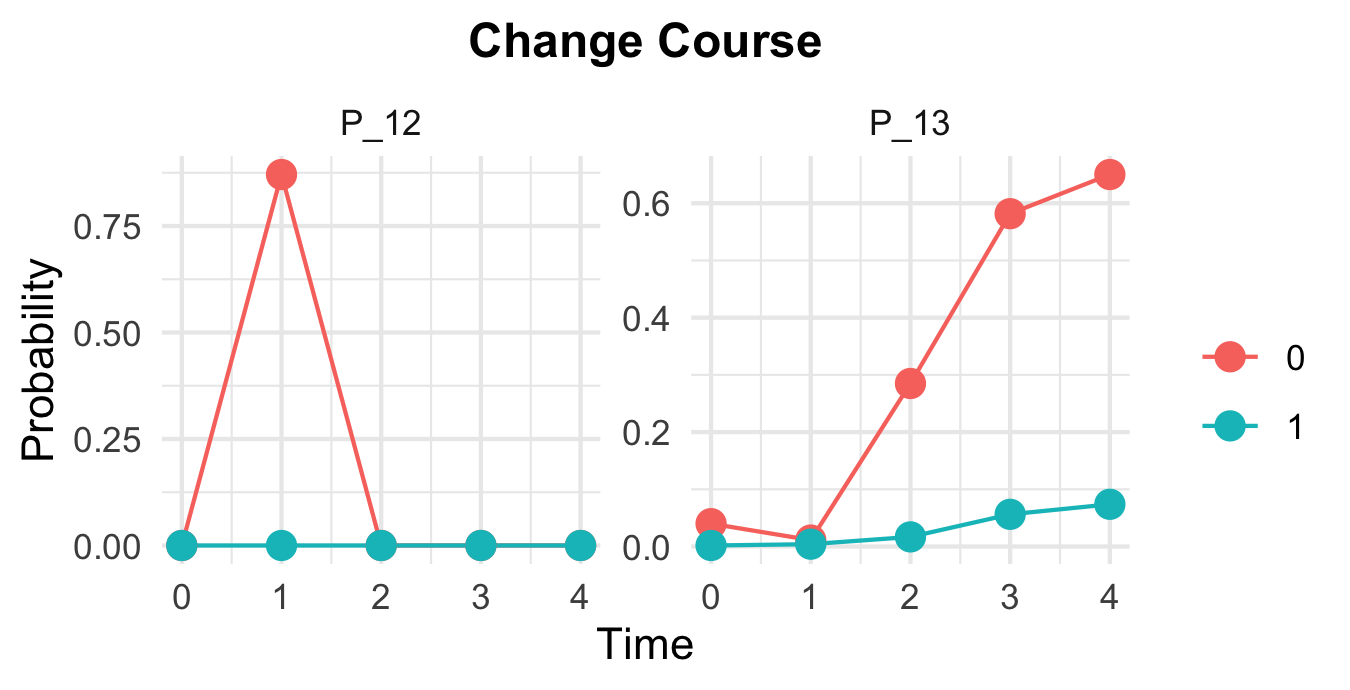}
        \caption{}
        \label{fig:change_course}
    \end{subfigure}
    \hfill
    \begin{subfigure}[b]{0.50\textwidth}
        \centering
        \includegraphics[width=\textwidth]{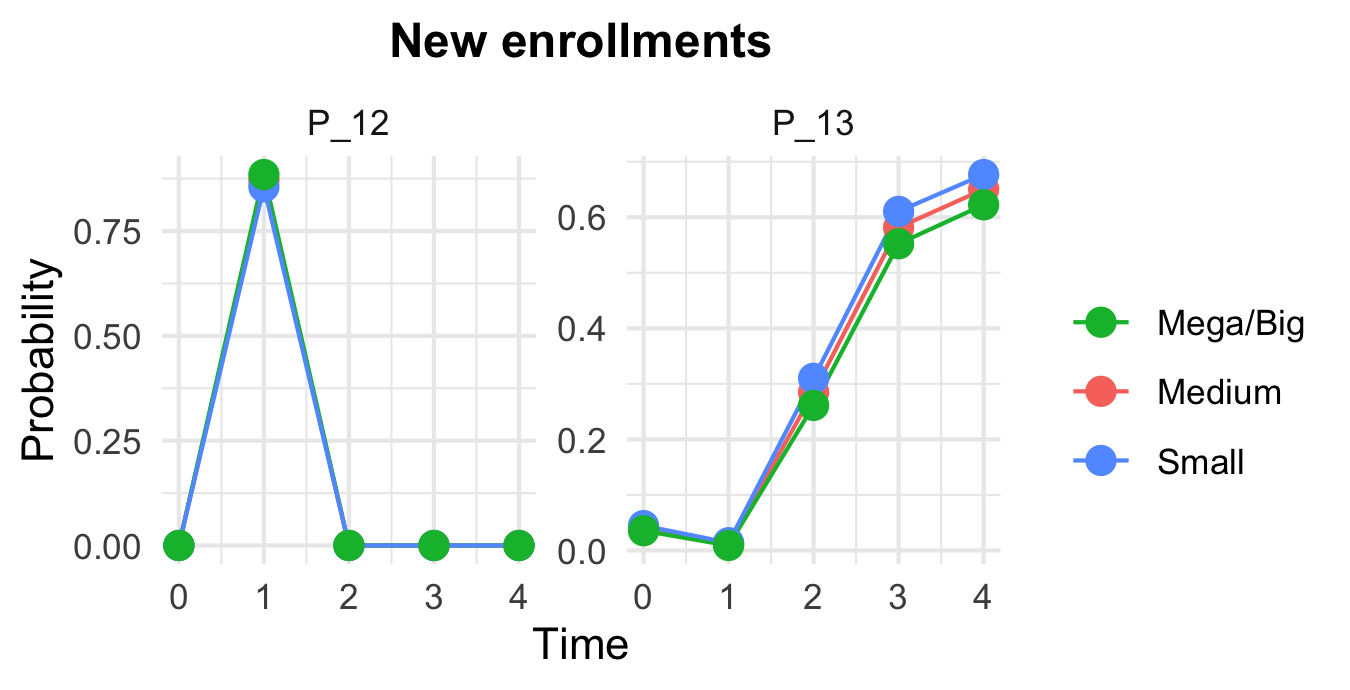}
        \caption{}
        \label{fig:new_enrollments}
    \end{subfigure}
    
    \vspace{0.5cm}
    
    \begin{subfigure}[b]{0.50\textwidth}
        \centering
        \includegraphics[width=\textwidth]{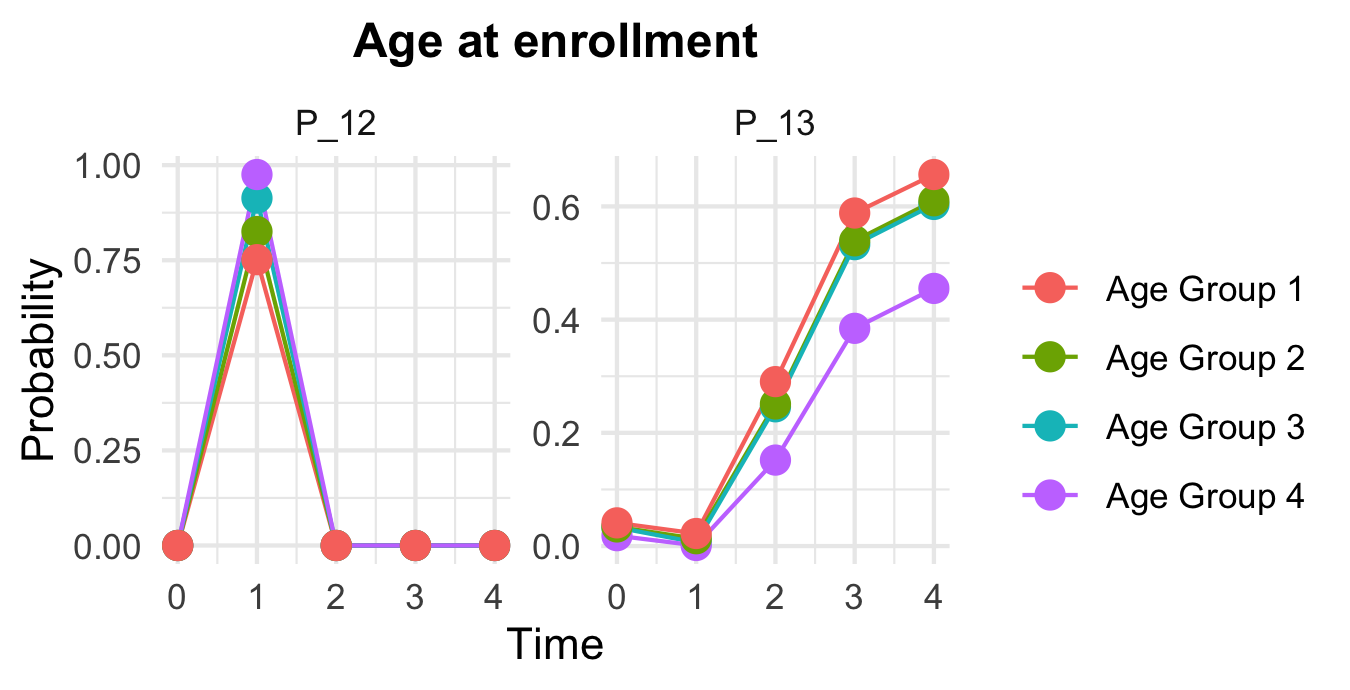}
        \caption{}
        \label{fig:age_enrollment}
    \end{subfigure}
    \hfill
    \begin{subfigure}[b]{0.50\textwidth}
        \centering
        \includegraphics[width=\textwidth]{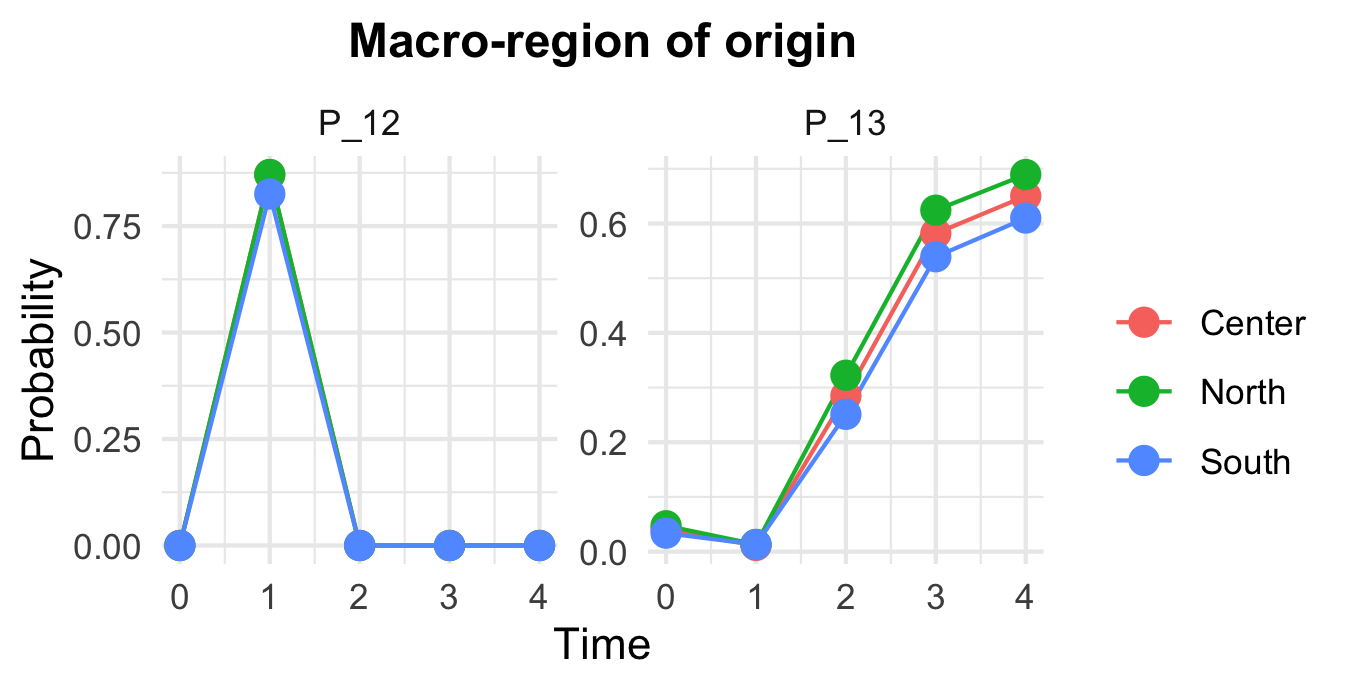}
        \caption{}
        \label{fig:macro_region}
    \end{subfigure}

    \caption{{Estimated transition probabilities to the stayer state $P_{12}$ and the mover state $P_{13}$ over time, stratified by selected covariates: (a) change of course (binary: yes (1), no(0)), university size (new enrollments: classified in small, medium and mega/big), age at enrollment (classified from youngest age-group 1 to oldest age-group 4), macro-region of origin (North, Center and South). For each covariate plot, other variables are held at their reference levels.}}
    \label{fig:combined_plots}
\end{figure}

The innovative aspect of the dynamic model is to also look at the trajectories of the probabilities $P_{12}$ and $P_{13}$ over time for some specific covariates profiles.
In Figure \ref{fig:combined_plots} the change in probabilities over time for different profiles is shown. Interesting to notice that for all the considered profiles at time 1 students achieve the highest probability of becoming a stayer: it is almost one for students belonging to the oldest age group (group 4) (Figure \ref{fig:combined_plots} (c)) and almost 0 for students that have performed at least one change of course (Figure \ref{fig:combined_plots} (a)). After time 1 the probability of becoming a stayer dramatically decreases for all profiles, while the probability of moving {(among those at risk)} strongly increases until time 3 and continues to increase at a lower rate until time 4. 
\begin{table}[h!]
\centering
\small
\begin{tabular}{l|cccc|cccc}
\multirow{2}{*}{{Parameter}} & \multicolumn{4}{c|}{\textbf{Static}} & \multicolumn{4}{c}{\textbf{Dynamic}} \\
 & Estimate & Lower CI & Upper CI & SD & Estimate & Lower CI & Upper CI & SD \\
\hline
$\alpha_0$ & -1.458 & -1.662 & -1.254 & 0.104 & 0.416 & -0.243 & 1.075 & 0.336 \\
$\alpha$[Age] & -0.496 & -0.669 & -0.323 & 0.088 & 0.032 & -0.081 & 0.145 & 0.058 \\
$\alpha$[Lyceum] & 0.455 & 0.340 & 0.570 & 0.059 & 12.200 & 12.198 & 12.202 & 0.001 \\
$\alpha$[Male] & -0.126 & -0.214 & -0.038 & 0.045 & 0.250 & -0.065 & 0.565 & 0.161 \\
$\alpha$[North] & -0.083 & -0.224 & 0.058 & 0.072 & -0.110 & -0.545 & 0.325 & 0.222 \\
$\alpha$[South] & 0.414 & -0.617 & 1.445 & 0.526 & -0.087 & -0.564 & 0.390 & 0.243 \\
$\beta_{{12}_0}$ & - & - & - & - & -20.000 & -20.068 & -19.932 & 0.035 \\
$\beta_{12}$[Age] & - & - & - & - & 0.430 & 0.243 & 0.617 & 0.095 \\
$\beta_{12}$[Lyceum] & - & - & - & - & 0.180 & -0.072 & 0.432 & 0.128 \\
$\beta_{12}$[Male] & - & - & - & - & 0.126 & 0.032 & 0.220 & 0.048 \\
$\beta_{12}$[North] & - & - & - & - & 0.018 & -0.109 & 0.145 & 0.065 \\
$\beta_{12}$[South] & - & - & - & - & -0.370 & -0.531 & -0.209 & 0.082 \\
$\beta_{{13}_0}$ & -1.961 & -2.225 & -1.697 & 0.135 & -3.180 & -3.391 & -2.969 & 0.108 \\
$\beta_{13}$[Age] & 0.004 & -0.099 & 0.107 & 0.053 & -0.200 & -0.302 & -0.098 & 0.052 \\
$\beta_{13}$[Lyceum] & 0.002 & -0.158 & 0.161 & 0.081 & -0.100 & -0.257 & 0.057 & 0.080 \\
$\beta_{13}$[Male] & 0.037 & -0.084 & 0.158 & 0.062 & -0.048 & -0.140 & 0.044 & 0.047 \\
$\beta_{13}$[North] & 0.279 & 0.092 & 0.466 & 0.095 & 0.178 & 0.058 & 0.298 & 0.061 \\
$\beta_{13}$[South] & -0.373 & -0.626 & -0.120 & 0.129 & -0.173 & -0.331 & -0.015 & 0.080 \\
$\gamma_{12}$[Size] & - & - & - & - & 0.080 & 0.031 & 0.129 & 0.025 \\
$\gamma_{12}$[Change] & - & - & - & - & -17.000 & -17.000 & -16.999 & 0.000 \\
$\gamma_{12}$[$t_1$] & - & - & - & - & 23.000 & 22.932 & 23.068 & 0.035 \\
$\gamma_{12}$[$t_2$] & - & - & - & - & 12.000 & 11.932 & 12.068 & 0.035 \\
$\gamma_{12}$[$t_3$] & - & - & - & - & -13.000 & -13.068 & -12.932 & 0.035 \\
$\gamma_{13}$[Size] & -0.140 & -0.195 & -0.085 & 0.028 & -0.080 & -0.129 & -0.031 & 0.025 \\
$\gamma_{13}$[Change] & -2.427 & -2.673 & -2.180 & 0.126 & -3.150 & -3.386 & -2.914 & 0.120 \\
$\gamma_{13}$[$t_1$] & -1.641 & -1.841 & -1.441 & 0.102 & 0.270 & 0.068 & 0.472 & 0.103 \\
$\gamma_{13}$[$t_2$] & 1.510 & 1.371 & 1.649 & 0.071 & 0.690 & 0.556 & 0.824 & 0.068 \\
$\gamma_{13}$[$t_3$] & -0.247 & -0.268 & -0.225 & 0.011 & -0.130 & -0.153 & -0.107 & 0.011 \\
\end{tabular}
\caption{Comparison of Static and Dynamic Model: maximum likelihood estimates, bootstrap confidence intervals and standard deviation. {Reference values: Non-Lyceum, Female, Center}}
\label{tab:bootstrap_comparison}
\end{table}
\begin{table}[ht]
\centering
\begin{tabular}{l|l|l}

{Variable} & {Level} & {Mean Probability} \\
\midrule
Age     & Low age    & 0.870 \\
Age     & Average age  & 0.769 \\
Age     & High age    & 0.687 \\
Sex     & Female                        & 0.832 \\
Sex     & Male                       & 0.849 \\
Diploma & Lyceum                        & 1.000 \\
Diploma & Non Lyceum                        & 0.613 \\
Macro-region   & North                       & 0.828 \\
Macro-region  & Center                        & 0.850 \\
Macro-region  & South                        & 0.849 \\

\end{tabular}
\caption{{Mean probability of initially being at risk of moving for selected levels of each covariate, computed while holding all other covariates at their baseline levels.}}
\label{tab:summary_probs}
\end{table}



\newpage
\section{Discussion}
\label{sec:conclusion}
This paper introduces a novel dynamic mover-stayer model for discrete time-to-event data, addressing an important limitation in existing models, namely, the assumption of a time-fixed stayer fraction. By allowing individuals initially at risk of experiencing an event (such as student mobility) to become stayers over time based on time-varying covariates, the model better reflects real-world dynamics, particularly in educational and social contexts.

The simulation studies demonstrate that the proposed dynamic model outperforms both static mover-stayer models and models that ignore stayers altogether. The results show that ignoring the time-dependent nature of the stayer status can lead to biased estimates, and that the dynamic model significantly reduces estimation errors as sample size increases. Importantly, the model still shows robustness under moderate sample sizes, although identifiability and numerical issues (such as separation and local maxima) may arise. 

The application to Italian university student mobility offers valuable insights. The analysis highlights strong effects of background factors, such as the type of high school diploma (Lyceum vs. non-Lyceum), age at enrollment, region of origin, and time-varying indicators like course changes and university size. Notably, students with Lyceum diplomas are more likely to be initially at risk of moving; older students and those enrolled in Southern universities are more likely to become stayers; students from Northern universities are more likely to preform a transition university-university. The probability of transitioning into a stayer status peaks after the first year, especially for students with stable academic trajectories (no change of course), highlighting the critical role of academic stability during the initial phase of university studies.

However, this work is not without limitations. In fact, from an implementation perspective, we encountered issues when using the EM algorithm, including non-monotonicity of the likelihood and convergence to local optima due to the non-convexity of the M-step. These issues have also been discussed in the literature on multinomial models with latent classes (e.g., \cite{durante2019nested}). Since the M-step requires numerical optimization, local maxima can compromise convergence. While adaptations of the EM algorithm could be considered, we argue that direct likelihood maximization, combined with multiple random initializations, is more straightforward and comparably effective in our setting. {With moderate sample size, a separation phenomenon could occur in practice leading to unrealistically large estimates.} A potential remedy could be the use of penalized likelihood methods to stabilize estimates, though this would introduce bias and raises challenges for valid inference post-model selection.
A key structural assumption in the model is that transitions from the at-risk state (state 1) to the stayer state (state 2) occur only once and are irreversible. Allowing for reversible transitions would make identifiability intractable when relying solely on exogenous covariates, due to the latent nature of these transitions.
We also note that the current framework assumes that time-dependent covariates are exogenous. This assumption simplifies the likelihood formulation but may be unrealistic in practice. For example, in our application, the covariate ‘change of course’ is arguably endogenous, as it may be influenced by the unobserved state of the student (mover vs. stayer). A more appropriate approach would involve joint modeling of the outcome and the covariates, as explored in the survival literature for internal time-varying covariates (e.g., \cite{barbieri2020joint}).
Moreover, we attempted to compute standard errors using the observed Fisher information matrix via numerical approximation of the Hessian, but this approach tended to underestimate the variance of parameter estimates. We recommend using a bootstrap procedure for variance estimation, which yielded more reliable results, especially for larger samples.

Future research could explore semiparametric extensions, though identifiability becomes more delicate. Another promising direction includes modeling reversible transitions (e.g., stayer to mover) in contexts where state switching is plausible and time-endogenous covariates. While such extensions would increase model complexity, they could better capture nuanced behavioral dynamics in evolving populations.

\section*{Acknowledgments}
MV conducted this research while affiliated with the University of Palermo (Italy) and acknowledges the financial support from the Italian Ministerial grant PRIN 2017 “From high school to job placement: micro-data life course analysis of university student mobility and its impact on the Italian North-South divide.”, n. 2017HBTK5P - CUP B78D19000180001

\section*{DATA ANALYSIS AND MANIPULATION }
The data used in this study have been processed in accordance with the RESEARCH PROTOCOL FOR THE
STUDY ``From high school to the job placement: analysis of university careers and university mobility from
Southern to Northern Italy" among the Ministry of University and Research, the Ministry of Education and
Merit, the University of Palermo as the lead institution, and the INVALSI Institute. The reference researcher is Massimo Attanasio.
\appendix
\section{EM algorithm}
\label{sec:EM}

Here we describe the EM algorithm for computation of the maximum likelihood estimator.  Let $B_i$ denote the initial susceptibility status of subject $i$: $B_i=1$ if at risk and $B_i=0$ for the initial stayers. Let $R_i$ denote the time of the transition `1$\rightarrow$2' for subjects initially at risk with $R_i=\infty$ if such transition does not occur. Note that for the censored observations both $B_i$ and $R_i$ are latent variables. 
The complete likelihood function, if $B_i$ and $R_i$ were observed is given by
\begin{equation*}
\begin{aligned}
    L_c(\bm{\alpha}, \bm{\beta}, \bm{\gamma}) &= 
    \prod_{i=1}^{n} 
    \Bigg\{ 
        (1 - \pi_i)^{1 - B_i} 
        \cdot \pi_i^{B_i} \left( \prod_{t=0}^{Y_i-1} P_{i,11}(t) P_{i,13}(Y_i) \right)^{ \Delta_i} \\
        &\quad\quad \times \left( \prod_{t=0}^{R_i-1} P_{i,11}(t) P_{i,12}({R_i}) \right)^{B_i (1 - \Delta_i) \bm{1}_{\{R_i \leq Y_i\}}} \left( \prod_{t=0}^{Y_i} P_{i,11}(t) \right)^{B_i (1 - \Delta_i) \bm{1}_{\{R_i > Y_i\}}}
    \Bigg\}
\end{aligned}
\label{eq:completelikelihood}
\end{equation*}
Hence we can write the log-likelihood function as
\begin{equation*}
\begin{aligned}
    l_c(\bm{\alpha}, \bm{\beta}, \bm{\gamma}) &= 
    \sum_{i=1}^{n} \left\{B_i\log\pi_i+(1-B_i)\log(1-\pi_i)\right\}+\sum_{i=1}^{n} \Delta_i\left\{\log P_{i,13}(Y_i)+\sum_{t=0}^{Y_i-1}\log P_{i,11}(t)\right\}\\
    &\quad+\sum_{i=1}^{n} (1-\Delta_i)B_i\sum_{t=0}^{Y_i}\left\{ 
        \bm{1}_{\{R_i=t\}} \log P_{i,12}(t) + \bm{1}_{\{R_i>t\}}  P_{i,11}(t) 
    \right\}.
\end{aligned}
\label{eq:completeloglikelihood}
\end{equation*}
Let $O_i=(Y_i,\Delta_i,\bm{x}_i,\bar{\bm{z}}_{Y_i})$ denote the observe data and $\bm\theta=(\bm{\alpha},\bm{\beta}_{12},\bm{\beta}_{13},\bm{\gamma}_{12},\bm{\gamma}_{13})$ the parameter estimated at the current iteration of the EM algorithm. For the E-step we compute
\[
\begin{aligned}
 W_i=\mathbf{E}\left[B_i\mid O_i,\bm\theta\right]=\Delta_i+(1-\Delta_i)\frac{\pi_1\prod_{t=0}^{Y_i}P_{i,11}(t)+\pi_i\sum_{r=0}^{Y_i}P_{i,12}(r)\prod_{t=0}^{r-1}P_{i,11}(t)}{1-\pi_i+\pi_1\prod_{t=0}^{Y_i}P_{i,11}(t)+\pi_i\sum_{r=0}^{Y_i}P_{i,12}(r)\prod_{t=0}^{r-1}P_{i,11}(t)},   
\end{aligned}
\]
\[
\begin{aligned}
   Q_i(r)&=\mathbf{E}\left[(1-\Delta_i)B_i \bm{1}_{\{R_i=t\}}\mid O_n,\bm\theta\right]\\
   &=\frac{(1-\Delta_i)\pi_iP_{i,12}(r)\prod_{t=0}^{r-1}P_{i,11}(t)}{1-\pi_i+\pi_1\prod_{t=0}^{Y_i}P_{i,11}(t)+\pi_i\sum_{r=0}^{Y_i}P_{i,12}(r)\prod_{t=0}^{r-1}P_{i,11}(t)}, \qquad r\leq Y_i,
\end{aligned}
\]
\[
\begin{aligned}
Q_i(\infty)&=\mathbf{E}\left[(1-\Delta_i)B_i \bm{1}_{\{R_i>Y_i\}}\mid O_n,\bm\theta\right]\\
&=\frac{(1-\Delta_i)\pi_i\prod_{t=0}^{Y_i}P_{i,11}(t)}{1-\pi_i+\pi_1\prod_{t=0}^{Y_i}P_{i,11}(t)+\pi_i\sum_{r=0}^{Y_i}P_{i,12}(r)\prod_{t=0}^{r-1}P_{i,11}(t)}.\qquad\qquad
\end{aligned}
\]
The M step then consists in maximizing the expected log-likelihood
\[
\begin{aligned}
    &\sum_{i=1}^{n} \left\{W_i\log\pi_i+(1-W_i)\log(1-\pi_i)\right\}+\sum_{i=1}^{n} \Delta_i\log P_{i,13}(Y_i)+\sum_{i=1}^{n}\sum_{t=0}^{Y_i} Q_i(t) \log P_{i,12}(t)\\
    &\qquad+\sum_{i=1}^{n} \sum_{t=0}^{Y_i}\left(Q_i(\infty)+\sum_{r=t+1}^{Y_i}Q_i(r)+\Delta_i\mathbf{1}_{\{t<Y_i\}}\right) \log P_{i,11}(t).
\end{aligned}
\]
The estimate for $\bm{\alpha}$ can be updated separately by maximizing the logistic log-likelihood $\sum_{i=1}^{n} \left\{W_i\log\pi_i+(1-W_i)\log(1-\pi_i)\right\}$. The other parameters are updated by maximizing the weighted log-likelihood of the multinomial logistic regression for the extended data constructed as follows. For each observed mover we create an observation corresponding to transition `1$\rightarrow$3' at time $Y_i$ with covariates $\bm{x},\bm{z}_{Y_i}$ and observations corresponding to transition `1$\rightarrow$1' at time $t$ with covariates $\bm{x},\bm{z}_{t}$ for each $t<Y_i$. For the censored subjects,  we create observations corresponding to transitions `1$\rightarrow$1' and `1$\rightarrow$2' at time $t$ with covariates $\bm{x},\bm{z}_{t}$ for each $t\leq Y_i$. Maximization can then be done using standard software for weighted MLE of multinomial logistic regression.
\section{Additional simulation results}
\label{sec:appendix}

Here we present the results of other two simulation settings described below. Setting 2 is similar to setting 1 (section \ref{sec:simulation}) but the proportion of initial stayers is lower, while setting 3 has a longer  follow-up of 10 time units.

\subsection{Simulation setting 2. }
We consider two independent baseline covariates: $x_i^1$ which has a standard normal distribution and $x_i^2$, which is a Bernoulli random variable with success probability $0.4$. The baseline mover-stayer status $B_0$ is generated as a Bernoulli variable with success probability $\pi(\bm\alpha;\bm{x}_i)$ given in \eqref{eq:curestatus} with $\bm\alpha=(2.3,0.5,-1)$.  We consider a maximum length follow-up $K=5$ and the event (moving) time $T$ can take values in $\{0,1,2,3,4\}$. At the beginning of each time unit we observe a bivariate vector of external time-dependent covariates $\bm{z}_{it}$ with components defined as follows. 
\[
\begin{cases}
 z^1_{i0}&=0 \\
 z^1_{it}&=z^1_{i,t-1}+{U^1_{it}}
\end{cases}\qquad \begin{cases}
    z^2_{i0}&\sim \text{Uniform}(\{1,2,3,4,5\}) \\
 z^2_{it}&=z^2_{i,t-1}+{U^2_{it}-1}
\end{cases}
\]
where {$(U^1_{t},U^2_{t})_{t\in\{0,\dots,4\}}$}, for $i=1,\dots,n $, are all independent and identically distributed, {$U^1_{t}\sim N(0.5,1)$ and $U^2_{t}\sim \text{Binom}(2,0.5)$.} At each time, we update the mover-stayer status $S_t$ according to the transition probabilities given in~\eqref{eqn:trans_prob} conditional on the covariate values with $\bm\beta_{12}=(-2,0.6,-0.1)$, $\bm\beta_{13}=(-1.5,-0.4,0.1)$, $\bm\gamma_{12}=(0.11,-0.2)$ and $\bm\gamma_{13}=(-0.5,0.3)$.  The censoring times are first generated from the continuous shifted exponential distribution with parameter 0.05 on $(1,\infty)$, and then are discretized and truncated at $4$. Table~\ref{tab:setting2} shows, for each time point, the percentage of observations in each of the three states, the percentage of observed movers and of censored observations. 

\begin{figure}
    \centering
    \includegraphics[width=0.8\textwidth]{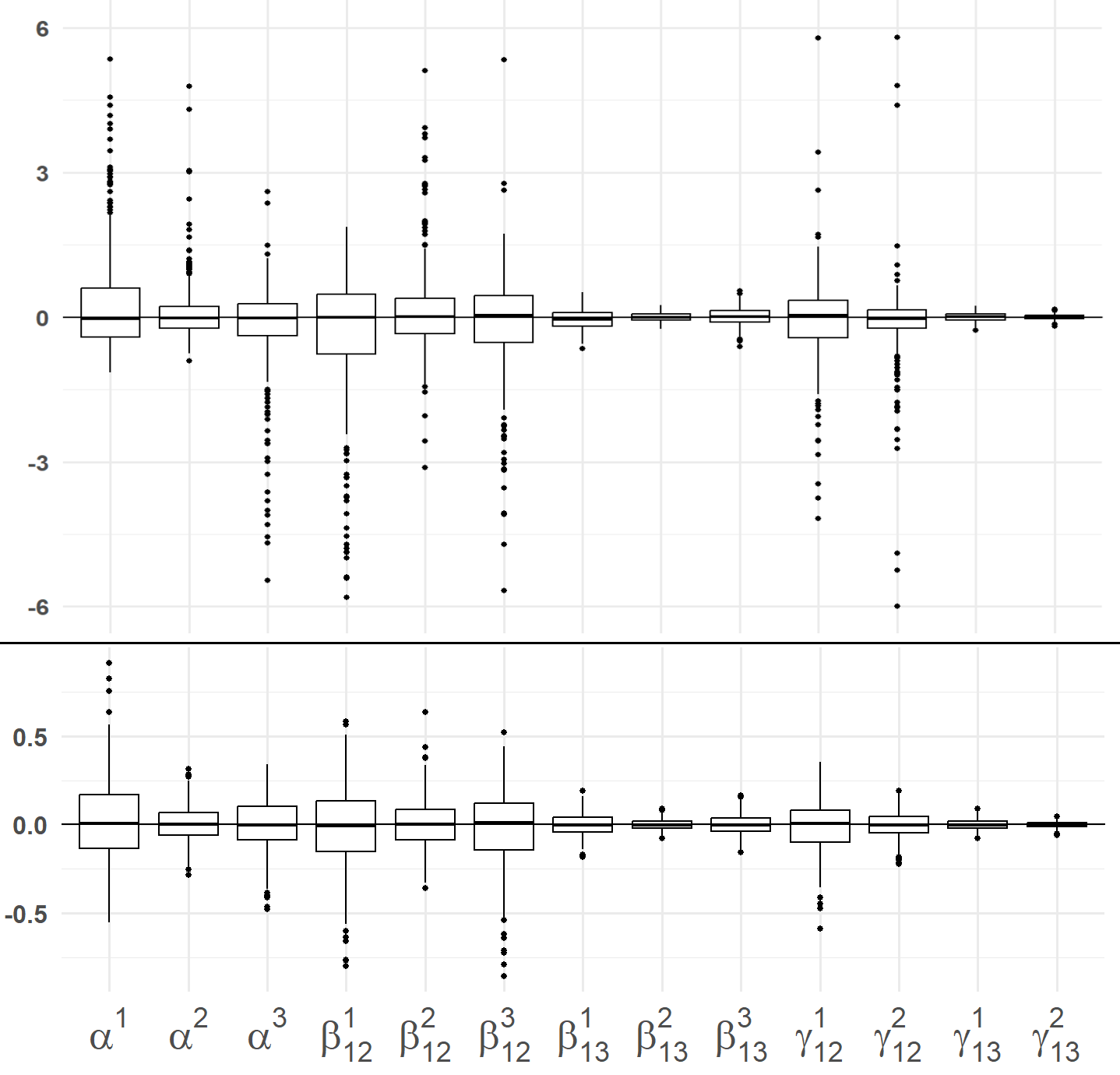}
    \caption{Boxplots of the parameter estimates centered at the true value for Setting 2 and sample size $n=1000$ (top), $n=10000$ (bottom).}
    \label{fig:est_2}
\end{figure}

\begin{figure}
    \centering
    \includegraphics[width=0.8\textwidth]{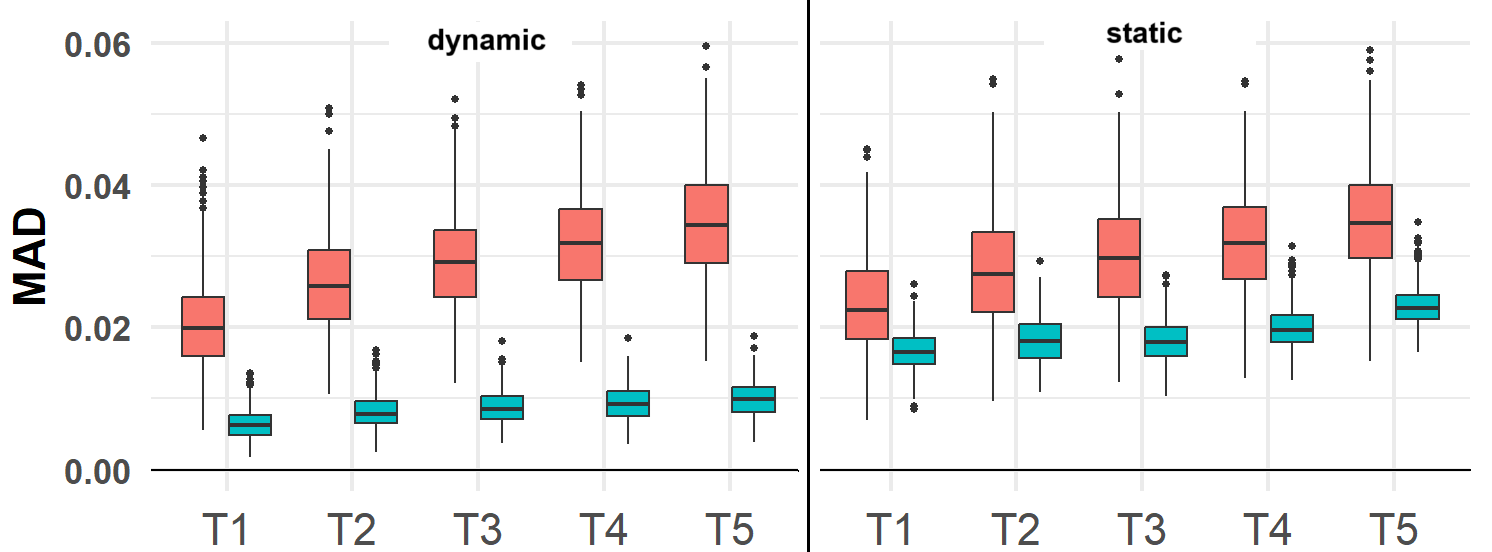}\\
    \includegraphics[width=0.45\textwidth]{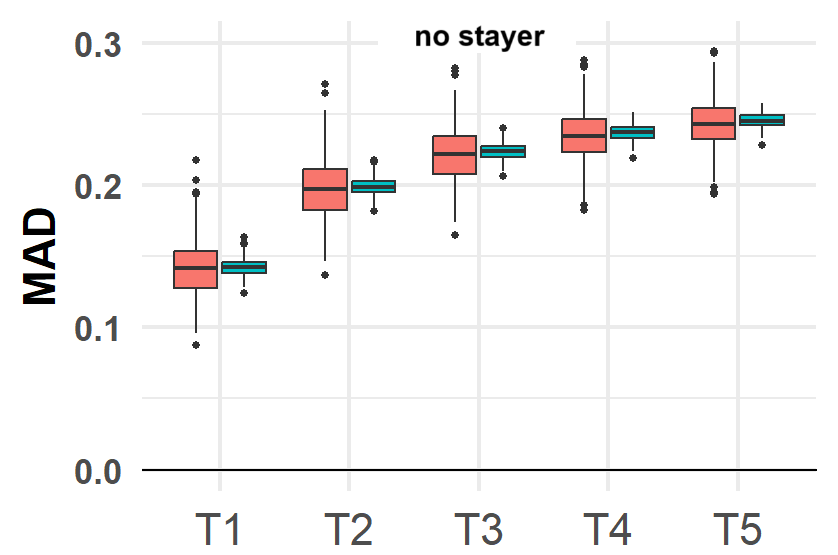}\\
    \caption{Boxplots of the average mean absolute deviation (MAD) for the cumulative probabilities of being a mover in Setting 2 and sample size $n=1000$ (red), $n=10000$ (green). Top left: dynamic model; Top right: static model; Bottom: model without stayers. }
    \label{fig:MAD_mover_2}
\end{figure}

\begin{figure}
    \centering
    \includegraphics[width=0.8\textwidth]{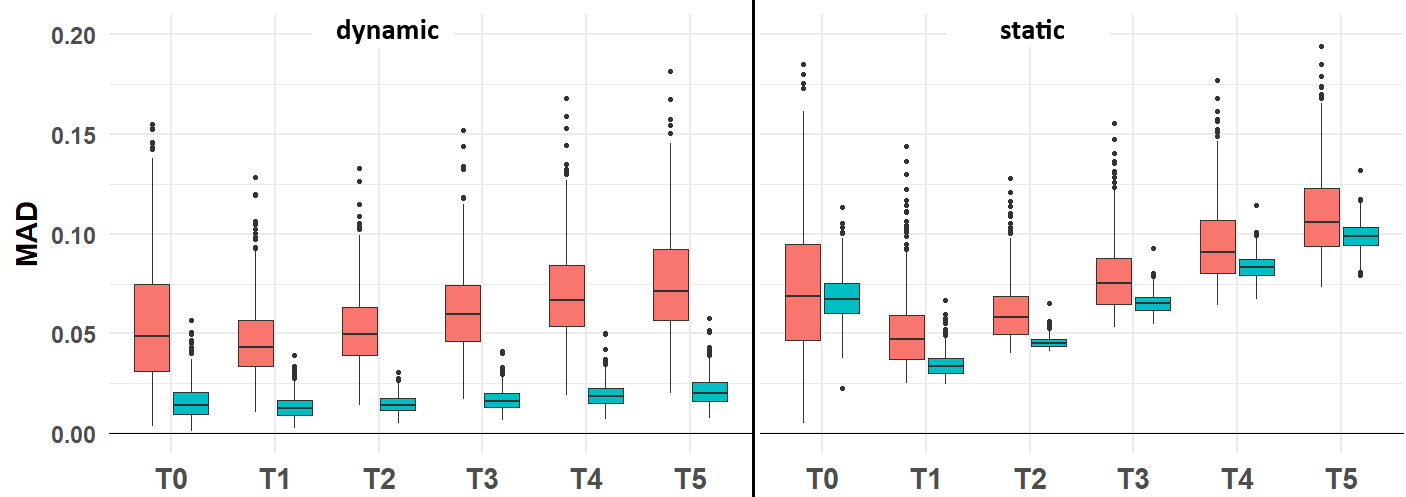}
    \caption{Boxplots of the average mean absolute deviation (MAD) for the cumulative probabilities of being a stayer in Setting 2 and sample size $n=1000$ (red), $n=10000$ (green). Left: dynamic model; Right: static model.}
    \label{fig:MAD_stayer_2}
\end{figure}

Boxplots of the parameter estimates over the 500 replications,
centered at the true values (i.e. $\hat{\bm{\theta}}_n-\bm\theta$), are shown in Figure~\ref{fig:est_2}. In
order to maintain good visibility of the figure, estimates with absolute value exceeding 6 are
not shown, which occurs in about $10-15\%$ of cases. The boxplots of the mean absolute deviation of the mover and
stayer probabilities, compared with the static model and the model without stayers, are shown in Figures~\ref{fig:MAD_mover_2}-\ref{fig:MAD_stayer_2} respectively. 

\begin{table}
    \centering
    \begin{tabular}{c|rrr|r|r}
    Time & in state 1 & in state 2& in state 3 & Obs. movers & Censored\\
    0 & 85\% & 15\%& 0\%  & 29.7\%  & 0\%  \\
    1 & 50\% & 20\% & 30\%  &  13.8\% & 3.4\% \\
    2 & 32\% & 24\%&44\%  &6.7\%   &2.6\%  \\
    3 & 22\%&26\% & 52\% &  3.4\% & 2.2\% \\
    4 &16\% &28\% &56\%  &  1.9\% & 36.3\% \\
    5 & 12\% & 30\%& 58\% &   &  \\
    \end{tabular}
    \caption{Percentage of observations in each state, of observed movers and of censored observations for each time point in setting 2.}
    \label{tab:setting2}
\end{table}


\subsection{Simulation setting 3.} 
 We consider two independent baseline covariates: $x_i^1$ which has a standard normal distribution and $x_i^2$, which is a Bernoulli random variable with success probability $0.4$. The baseline mover-stayer status $B_0$ is generated as a Bernoulli variable with success probability $\pi(\bm\alpha;\bm{x}_i)$ given in \eqref{eq:curestatus} with $\bm\alpha=(0.8,0.5,-1)$.  We consider a maximum length follow-up $K=10$ and the event (moving) time $T$ can take values in $\{0,1,\dots,9\}$. At the beginning of each time unit, we observe a bivariate vector of external time-dependent covariates $\bm{z}_{it}$ with components defined as follows. 
\[
\begin{cases}
 z^1_{i0}&=0 \\
 z^1_{it}&=z^1_{i,t-1}+{U^1_{it}}
\end{cases}\qquad \begin{cases}
    z^2_{i0}&\sim \text{Uniform}(\{1,2,3,4,5\}) \\
 z^2_{it}&=z^2_{i,t-1}+{U^2_{it}-1}
\end{cases}
\]
where {$(U^1_{t},U^2_{t})_{t\in\{0,\dots,4\}}$}, for $i=1,\dots, n$, are all independent and identically distributed, {$U^1_{t}\sim N(0.5,1)$ and $U^2_{t}\sim \text{Binom}(2,0.5)$}. At each time, we update the mover-stayer status $S_t$ according to the transition probabilities given in~\eqref{eqn:trans_prob} conditional on the covariate values with $\bm\beta_{12}=(-1,0.6,-0.1)$, $\bm\beta_{13}=(-2,-0.1,0.3)$, $\bm\gamma_{12}=(0.2,-0.2)$ and $\bm\gamma_{13}=(-0.1,0.1)$.  The censoring times are first generated from the continuous shifted exponential distribution with parameter 0.03 on $(1,\infty)$, and then are discretized and truncated at $9$. Table~\ref{tab:setting3} shows, for each time point, the percentage of observations in each of the three states, the percentage of observed movers and of censored observations.

\begin{figure}
    \centering
    \includegraphics[width=0.8\textwidth]{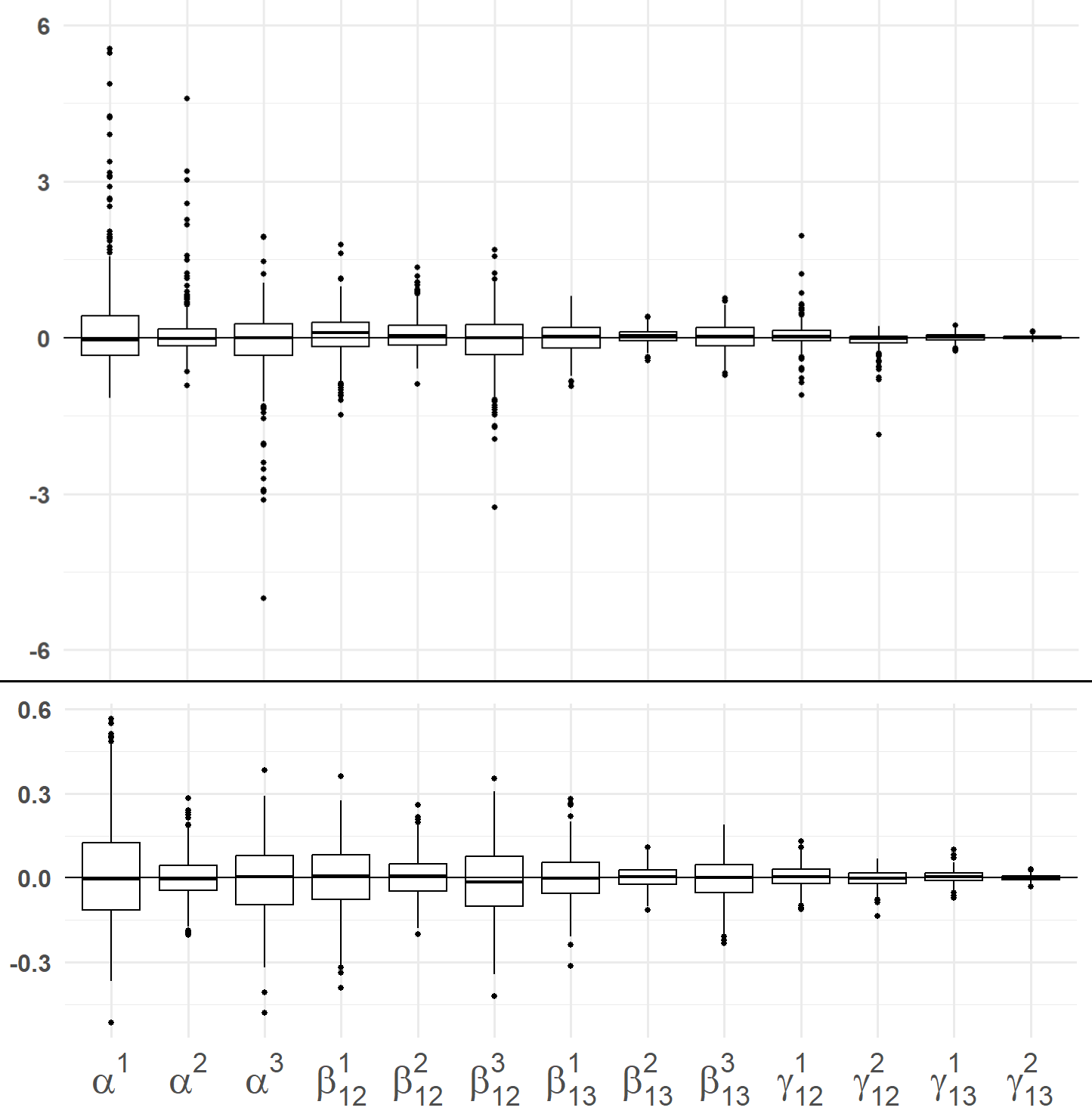}
    \caption{Boxplots of the parameter estimates centered at the true value for Setting 3 and sample size $n=1000$ (top), $n=10000$ (bottom).}
    \label{fig:est_3}
\end{figure}

\begin{figure}
    \centering
    \includegraphics[width=\textwidth]{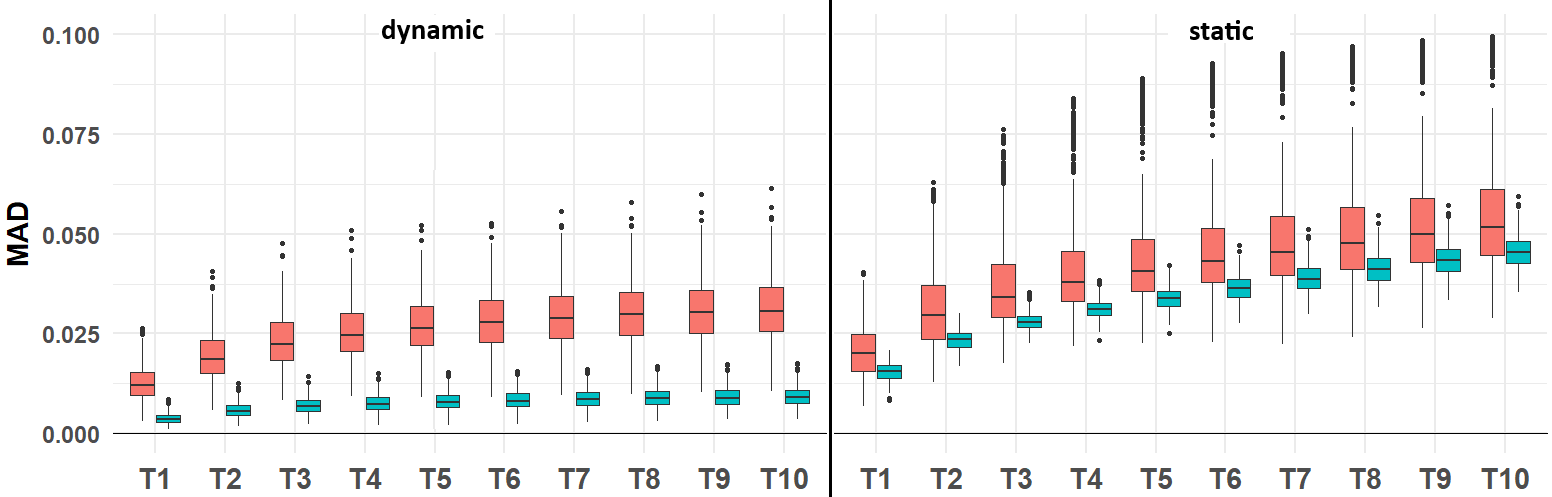}\\
     \includegraphics[width=0.5\textwidth]{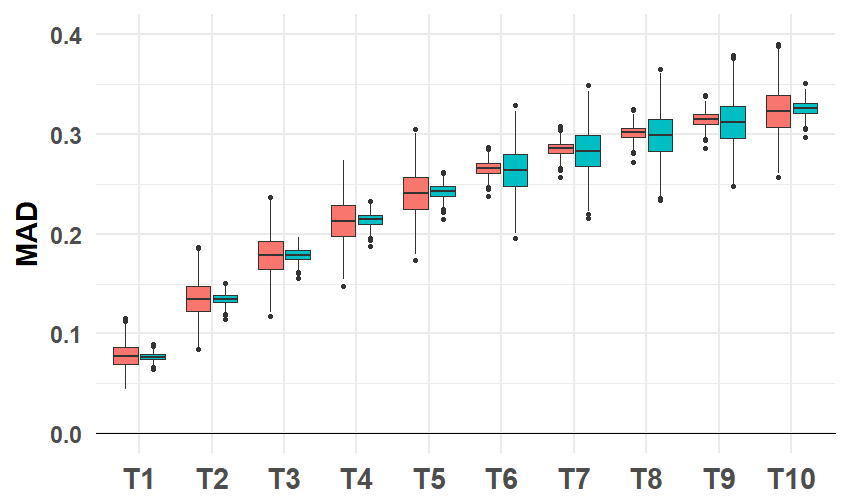}
    \caption{Boxplots of the average mean absolute deviation (MAD) for the cumulative probabilities of being a mover in Setting 3 and sample size $n=1000$ (red), $n=10000$ (green). Top left: dynamic model; Top right: static model; Bottom: model without stayers. }
    \label{fig:MAD_mover_3}
\end{figure}

\begin{figure}
    \centering
    \includegraphics[width=\textwidth]{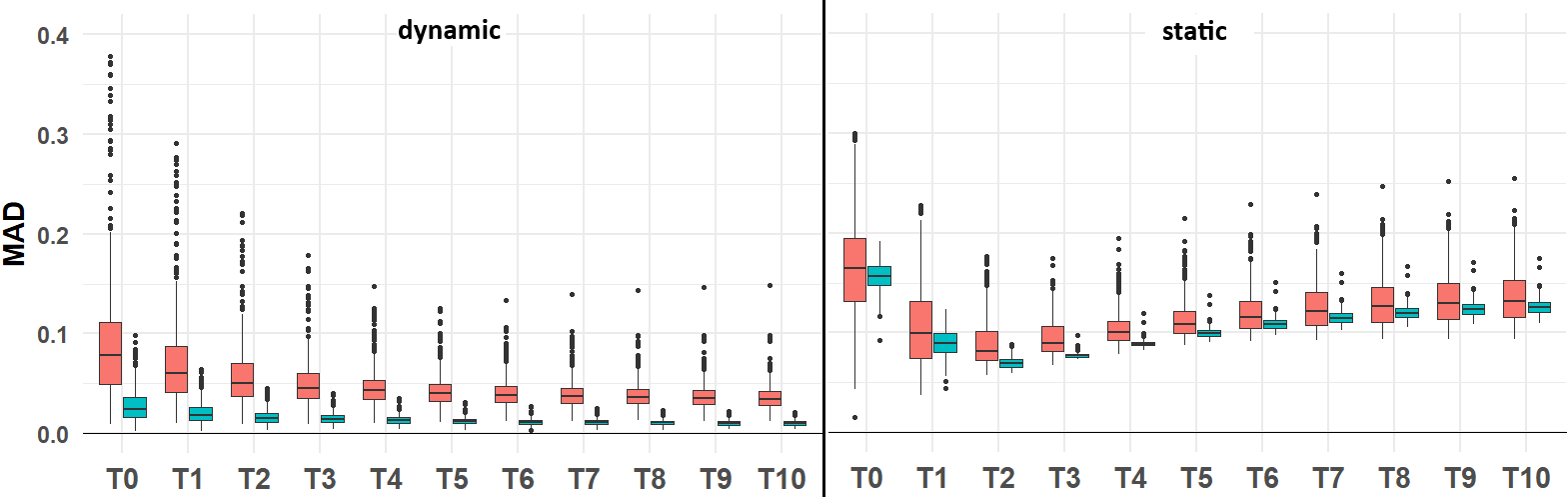}
    \caption{Boxplots of the average mean absolute deviation (MAD) for the cumulative probabilities of being a stayer in Setting 3 and sample size $n=1000$ (red), $n=10000$ (green). Left: dynamic model; Right: static model.}
    \label{fig:MAD_stayer_3}
\end{figure}

Boxplots of the parameter estimates over the 500 replications,
centered at the true values (i.e. $\hat{\bm{\theta}}_n-\bm\theta$), are shown in Figure~\ref{fig:est_3}. In
order to maintain good visibility of the figure, estimates with absolute value exceeding 6 are
not shown, which occurs in about $1-2\%$ of cases. The boxplots of the mean absolute deviation of the mover and
stayer probabilities, compared with the static model and the model without stayers, are shown in Figures~\ref{fig:MAD_mover_3}-\ref{fig:MAD_stayer_3} respectively. 
\begin{table}
    \centering
    \begin{tabular}{c|rrr|r|r}
    Time & in state 1 & in state 2& in state 3 & Obs. movers & Censored\\
    0 & 59\% & 41\%& 0\%  & 10.9\%  & 0\%  \\
    1 & 41\% & 48\% & 11\%  &  7.3\% & 2.4\% \\
    2 & 28\% & 53\%&18\%  &4.7\%   &2.2\%  \\
    3 & 20\%&57\% & 23\% &  3.1\% & 2.0\% \\
    4 &14\% &60\% &26\%  &  2.0\% & 1.9\% \\
    5 & 9\% & 62\%& 29\% &  1.3\% & 1.8\% \\
    6 & 6\% & 64\%& 30\%  & 0.8\%  & 1.8\%  \\
    7 & 4\% & 65\% & 31\%  &  0.5\% & 1.7\% \\
    8 & 3\% & 65\%&32\%  &0.3\%   &1.7\%  \\
    9 & 2\%&66\% & 32\% &  0.2\% & 53.3\% \\
    10 &1\% &66\% &32\%  &   &\\
    \end{tabular}
    \caption{Percentage (rounded) of observations in each state, of observed movers and of censored observations for each time point in setting 3.}
    \label{tab:setting3}
\end{table}

\newpage
\printbibliography

\end{document}